\documentstyle[preprint,aps]{revtex}
\draft
\begin{document}
\addtolength{\voffset}{1.1in}
            
\title{\bf Phenomenology of neutral heavy leptons}
\author{Pat Kalyniak  and I. Melo}
\address{Ottawa-Carleton Institute for Physics\\
Department of Physics, Carleton University\\
1125 Colonel By Drive,
Ottawa, Ontario, Canada K1S 5B6}
                                                      
\maketitle

\begin{abstract}
{We continue our previous work on the flavour-conserving
leptonic decays of the $Z$ boson with neutral heavy leptons (NHL's) in the loops
by considering box, vertex, and self-energy diagrams for the muon decay. By
inclusion of these loops (they contribute to the input parameter
$M_{W}$) we can probe the full parameter space
spanned by the so-called flavour conserving mixing parameters $ee_{mix},
\mu\mu_{mix}, \tau\tau_{mix}$.
We show that only two diagrams from each class (box, vertex and self-energy)
are important; further, after renormalization only two box diagrams 'survive' 
as dominant. We compare the results of our analysis with the existing work
in this field  and conclude that flavour-conserving decays have certain
advantages over traditionally considered flavour-violating ones.}
\end{abstract}

\pacs{PACS number(s): 14.60.St, 12.15.Ff, 13.35.Bv}

\newpage
\section{Introduction}

We have previously considered \cite{melo} a simple extension of
the standard model (SM) with an enriched neutral fermion spectrum consisting of a
massless neutrino and a Dirac neutral heavy lepton (NHL) associated with each
generation \cite{vallemo,bernabeu1,wolfe}. Several parameters can be used to characterize the model:
`flavour-conserving' mixing parameters $ee_{mix}, \mu\mu_{mix}, \tau\tau_{mix}$;
`flavour-violating' mixing parameters $e\mu_{mix}, e\tau_{mix}, \mu\tau_{mix}$
and the mass scale  $M_{N}$ of NHL's (assuming three degenerate NHL's). We
considered the effect, via these parameters, of NHL's on flavour-conserving
$Z$ boson decays to charged leptons and on the $W$ boson mass, $M_{W}$. However,
in our earlier work, we neglected all mixing parameters except $\tau\tau_{mix}$,
which is the least well constrained. Here we generalize our analysis 
by considering the case of arbitrary mixings
$ee_{mix}, \mu\mu_{mix}$ and $\tau\tau_{mix}$. Our previous neglect of
$ee_{mix}$ and  $\mu\mu_{mix}$ allowed us to also neglect a number of
contributions to the muon decay corrections which feed into $M_{W}$ as
an input parameter. Including these couplings, non-SM box, vertex and self-energy
diagrams contributing to the muon decay (see Figs.~\ref{boxmuon}, \ref{vermuon},
\ref{selfmuon}) may become important for the calculation of  $M_{W}$. In our
previous paper \cite{melo}, as a result of the assumption $ee_{mix} =
\mu\mu_{mix} = 0$, only oblique corrections (corrections to the W propagator)
had to be considered. Here we consider the full set of corrections.
Still, we assume here vanishing flavour-violating mixing parameters:
$e\mu_{mix}, e\tau_{mix}, \mu\tau_{mix} = 0$.
These parameters, if nonzero, lead to further complications, which in general
require,
as argued in a recent work \cite{kniehl}, the renormalization of the
mixing matrix. This is an interesting topic by itself; nevertheless, it is not
crucial for our considerations. We note that the smallness of $e\mu_{mix}$
is confirmed by experiment \cite{pdb,Ng1,Ng2,ggjv,Ilakovac,Jarlskog}.

The inclusion of the arbitrary flavour-conserving mixing parameters completes our studies of the NHL's impact on the processes considered here. We compare our constraints on the parameters of the model with those coming from the traditionally favoured flavour-violating processes, such as $\mu \rightarrow e \gamma$, $\tau \rightarrow e e e$, $Z \rightarrow e \mu$, etc \cite{bernabeu1,Ng1,Ng2,ggjv,Ilakovac,Jarlskog,Valle2,Korner,pilaftsis1}. We find that the processes we consider have certain advantages over the latter ones.

This paper is organized as follows. In Sec. II below, we briefly review
a superstring-inspired $SU(2)_{L} \times U(1)_{Y}$ model of neutrino mass and the constraints on the mixings and masses of the model.
In Sec. III, we present the additional
muon decay corrections, identifying which contributions are important.
Ultimately, our earlier results can primarily be improved by the tree level modification
of the vertex by mixing factors. In the limit of large $M_{N}$, only two box
diagrams finally contribute but these are numerically only marginally important.
Given the muon decay corrections, we also present the one-loop modification of
the constraint on $\tau \tau_{mix}$. In Sec. IV, we consider more generally the
work done in this field. We contrast the sensitivity to the presence of NHL's
in flavour-violating processes with the results for flavour-conserving
processes. We include a calculation of the flavour-violating leptonic decays of
the $Z$ boson in our model. We summarize in Sec. V.

\section{A superstring-inspired $SU(2)_{L} \times U(1)_{Y}$ model of neutrino mass}

Here we briefly describe the model of neutrino mass which we consider. For more details, we refer the reader to the original papers \cite{vallemo,bernabeu1,wolfe} or our previous work \cite{melo}. The model extends the neutral fermion sector of the SM by two new weak isosinglet neutrino fields ($n_{R}, S_{L}$) per generation. With total lepton number conservation imposed, the mass matrix is given by
\begin{eqnarray}
\label{ourmatrix}
-{\cal L}_{mass} & = & \frac{1}{2}{\cal M} \; = \;
 \frac{1}{2}\left(\overline{\nu_{L}}\;\overline{n_{L}^{c}}\;\overline{S_{L}}
\right)
\left( \begin{array}{lll}
                       0     & D & 0      \\
                       D^{T} & 0 & M^{T}  \\
                       0     & M & 0
               \end{array} \right)
\left( \begin{array}{c}
                       \nu_{R}^{c} \\
                       n_{R}      \\
                       S^{c}_{R}
               \end{array} \right) + h.c..
\end{eqnarray}
Each $\nu_{L},n_{R},S_{L}$ represents a collection of three fields,
one for each family. $D$ and $M$ are $3 \times 3$
matrices. The diagonalization of the mass matrix yields three massless neutrinos ($\nu_{i}$) along with three Dirac NHL's ($N_{a}$) of mass $M_{N} \sim M$. The weak interaction eigenstates ($\nu_{l}, l=e,\mu,\tau$) are related to the six mass eigenstates via a $3 \times 6$ mixing matrix $K \equiv (K_{L}, K_{H})$:
\begin{eqnarray}
\label{alter}
\nu_{l} & = & \sum_{i=1,2,3}\big(K_{L}\big)_{li} \nu_{i_{L}}
+ \sum_{a=4,5,6} \big(K_{H}\big)_{la} N_{a_{L}}.
\end{eqnarray}
The mixing factor which typically governs flavour-conserving
processes, $ll_{mix}$,
is given by
\begin{eqnarray}
ll_{mix} & = & \sum_{a=4,5,6} \big(K_{H}\big)_{la}
\big(K_{H}^{\dagger}\big)_{al}\;\; ;
\makebox[.5in] [c] { } l= e, \mu, \tau
\end{eqnarray}
and the flavour-violating mixing factor   $l{l^{'}}_{mix}$
is defined as
\begin{eqnarray}
l{l^{'}}_{mix} & = & \sum_{a=4,5,6} \big(K_{H}\big)_{la}
\big(K_{H}^{\dagger}\big)_{al^{'}}\;\; ;
\makebox[.5in] [c] { } l,l^{'} = e, \mu, \tau,
\makebox[.2in] [c] { }l \neq l^{'}.
\end{eqnarray}
Further, the following important inequality holds
\begin{eqnarray}
\label{ineq}
|{l{l^{'}}_{mix}}|^2 & \leq & {ll}_{mix}\:\:{l^{'}l^{'}}_{mix},
\makebox[.5in] [c] { } l \neq l^{'}.
\end{eqnarray}
This implies  that one might observe nonstandard effects in flavour-conserving
processes even if they are absent in flavour-violating processes.

We note here existing constraints on the parameter space of the model.
{\it Indirect constraints} on the flavour-conserving mixing parameters $ee_{mix}, \mu\mu_{mix},
\tau\tau_{mix}$ have been obtained from a global analysis of results including lepton
universality measurements, CKM matrix unitarity tests, $W$ mass
measurement, and LEP I measurements. These
constraints arise primarily at tree level due to the modification of
couplings from those of the SM. Nardi et al \cite{Nardi} have found the following upper limits:
\begin{eqnarray}
\label{limits1}
      ee_{mix} & \leq & 0.0071 \nonumber \\
      \mu\mu_{mix} & \leq  & 0.0014  \\
      \tau\tau_{mix} & \leq & 0.033 \nonumber
\end{eqnarray}
Since the limit on the parameter $\tau\tau_{mix}$ plays
(as the least stringent one so far) the most important role in our analysis,
we will pay further attention to its source. The $\mu - \tau$ universality test is based on the $\tau$ leptonic decays compared
to the $\mu$ leptonic decays, with the result given as the ratio of
the couplings of $\tau$ and $\mu$ to the $W$ boson, $g_{\tau}/g_{\mu}$.
The tree level ratio is found
from 
\begin{eqnarray}
\label{modif}
\frac{\Gamma(\tau \rightarrow e \nu \nu) /
\Gamma^{SM}(\tau \rightarrow e \nu \nu)}
{\Gamma(\mu \rightarrow e \nu \nu) /
\Gamma^{SM}(\mu \rightarrow e \nu \nu)}
& = & \Big(\frac{g_{\tau}}{g_{\mu}}\Big)^{2} \; = \;
\frac{1-\tau\tau_{mix}}{1-\mu\mu_{mix}}.
\end{eqnarray}
This measurement has undergone substantial improvement over the last 
while. With the most recent result $g_{\tau}/g_{\mu} = 0.9994 \pm 0.0028$
\cite{moriond}, the constraint on $\tau\tau_{mix}$ is improved from its previous value of 0.033 by a factor of about three.
To reflect this improvement we present most of our results either for the values of $\tau\tau_{mix}$ ranging
from 0.033 to 0.01, or in a general form with $\tau\tau_{mix}$ as a variable.
In a few cases (e.g. when quoting results of others on flavour-violating processes), however, we only use $\tau\tau_{mix}=0.033$. Finally, we note that these
indirect limits depend very weakly on $M_{N}$; this point will be illustrated
 at the end of Sec. III. 

Since NHL's have not been directly observed in the Z decay $Z \rightarrow N \nu$, we focus on NHL masses $M_{N} \gg M_{Z}, M_{W}, M_{H}$. These can be probed indirectly via {\it loop effects}
in either flavour-violating or flavour-conserving processes. 
As argued in our previous work \cite{melo}, only in this case are the contributions of NHL's via loops possibly significant, due to the violation of the Appelquist-Carrazzone decoupling theorem \cite{ac}. 
Analogous to the behaviour of the top quark loop contributions in the SM,
quadratic nondecoupling (amplitudes $\sim M_{N}^{2}$) often results here. 

\section{NHL contributions to muon decay}
                                 
\subsection{Box diagrams}
\label{secbox}

We first consider the box diagrams contributing to the muon decay, as
depicted in Fig.~\ref{boxmuon}. Diagrams of Fig.~\ref{boxmuon}b,f,g each come in
two varieties, with either massless neutrinos or NHL's in the loop. All diagrams
without NHL's are similar to their SM counterparts; the only slight difference
comes from the mixing factors in vertices
(such as $1-ll_{mix}$, see Appendix A).  All the box graphs are finite.

%\begin{figure}[hbtp]
%\begin{center}
%\setlength{\unitlength}{1in}
%\begin{picture}(6,3)
%\put(.3,+0.325){\mbox{\epsfxsize=5.0in\epsffile{p2fig1.eps}}}
%\end{picture}
%\end{center}
%\caption{Box diagrams for muon decay}
%\label{boxmuon}
%\end{figure}

The results of our computation of the diagrams of Fig.~\ref{boxmuon}a-g are given
in Appendix A. (The QED box amplitude of Fig.~\ref{boxmuon}h, 
${\cal M}_{\gamma e W \mu}$, is given in Ref.~\cite{key6}.) 
The dominant nonstandard
contribution in the limit of $M_{N} \gg M_{Z}, M_{W}, M_{H}$ comes from just two graphs depicted
in Fig.~\ref{boxmuon}e, one with Higgs H and one with neutral unphysical Higgs
$\chi$. To show that these two graphs are dominant we could take the large
$M_{N}$ limit of the exact results given in Appendix A. We would find that only
these two graphs exhibit quadratic nondecoupling, i.e., quadratic overall
 dependence on $M_{N}$.
The remaining graphs with NHL's are either constant in the large $M_{N}$ limit,
or decouple as $1/M_{N}^{2}$. However, here we prefer a more intuitive approach 
based on dimensional analysis considerations and power counting.

The amplitude for the diagram with the Higgs boson $H$ (Fig.~\ref{boxmuon}e)
is given by
(we sum over NHL's $N_{a}, N_{b}$ with $M_{N_{a}} = M_{N_{b}} = M_{N}$ and
neglect external momenta in the internal propagators)
\begin{eqnarray}
\label{amp1}
{\cal M}_{\phi N H N} & = & \sum_{a,b} \int \frac{d k^{4}}{(2\pi)^{4}}
\overline{u}_{\nu_{\mu}}\frac{-i
g_{2}}{2}\frac{M_{N}}{M_{W}}\big(K_{L}^{\dagger}K_{H}\big)_{ia}
\frac{1+\gamma_{5}}{2}\frac{i}{\not k - M_{N}}\frac{ig_{2}}{2\sqrt{2}}
\frac{M_{N}}{M_{W}}\big(K_{H}^{\dagger}\big)_{a\mu} \nonumber \\
& \times &
(1-\gamma_{5})u_{\mu}\;\;\overline{v}_{e}\frac{ig_{2}}{2\sqrt{2}}
\frac{M_{N}}{M_{W}}\big(K_{H}\big)_{eb}(1+\gamma_{5})
\frac{i}{\not k - M_{N}}\frac{-i
g_{2}}{2}\frac{M_{N}}{M_{W}}\big(K_{H}^{\dagger}K_{L}\big)_{bj} \nonumber \\
& \times  &
\frac{1-\gamma_{5}}{2}v_{\nu_{e}}\frac{i}{k^{2}-M_{W}^{2}}
\frac{i}{k^{2}-M_{H}^{2}}.
\end{eqnarray}
Various mixing factors can be collected as
\begin{eqnarray}
k_{mix} \equiv 
\big(K_{L}^{\dagger}K_{H}\big)_{ia}\big(K_{H}^{\dagger}\big)_{a\mu}
\big(K_{H}\big)_{eb}\big(K_{H}^{\dagger}K_{L}\big)_{bj} 
& = & {\big(K_{L}^{\dagger}\big)_{i\mu}
\big(K_{L}\big)_{ej}}ee_{mix}\mu\mu_{mix},\;\;\;
\end{eqnarray}
where we used $e\mu_{mix} = \mu\tau_{mix} = 0$.
%***************************************
Neglecting some constant factors which we will restore later, we get
\begin{eqnarray}
{\cal M}_{\phi N H N} & \sim & M_{N}^{4} 
\int \frac{d^{4}k}{{(2\pi)}^{4}}
\frac{k^{2}}{(k^{2}-M_{N}^{2})^{2}(k^{2}-M_{W}^{2})(k^{2}-M_{H}^{2})}.
\end{eqnarray}
Note that the Lorentz structure of the amplitude is such that NHL propagators
$\frac{i}{{\not k}-M_{N}}$ contribute as $\frac{i {\not k}}{{k}^{2}-M_{N}^{2}}$
rather than $\frac{i M_{N}}{{k}^{2}-M_{N}^{2}}$.
In the limit of large $M_{N}$ we can neglect all masses and momenta except 
$M_{N}$, obtaining
\begin{eqnarray}
{\cal M}_{\phi N H N} & \sim & M_{N}^{4} 
\int \frac{d^{4}k}{{(2\pi)}^{4}}
\frac{1}{(k^{2}-M_{N}^{2})^{2} k^{2}}.
\end{eqnarray}
The integral is expected to be of the form $(M_{N})^{p}$; power counting 
yields
$p = -2$, so indeed the amplitude depends quadratically on $M_{N}$:
\begin{eqnarray}
{\cal M}_{\phi N H N} & \sim & M_{N}^{4} M_{N}^{-2} \; = \; M_{N}^{2}.
\end{eqnarray}
We can further improve our estimate by restoring the constants collected
from Eq.~\ref{amp1},
\begin{eqnarray}
\label{pnhn}
{\cal M}_{\phi N H N} & = & c {\cal M}_{tree} \frac{\alpha}{s_{W}^{2}}
\frac{M_{N}^{2}}{M_{W}^{2}} ee_{mix} \mu\mu_{mix},
\end{eqnarray}
where 
\begin{eqnarray}
{\cal M}_{tree} & = & - \frac{i g_{2}^{2}}{8 M_{W}^{2}}
\big[\overline{u}_{\nu_{\mu}}
(1+\gamma_{5})\gamma_{\alpha}u_{\mu}\big]
\big[\overline{v}_{e}(1+\gamma_{5}) \gamma^{\alpha} v_{\nu_{e}}\big]
{\big(K_{L}^{\dagger}\big)_{i\mu} \big(K_{L}\big)_{ej}},
\end{eqnarray}
is the tree-level amplitude. The remaining numerical factor $c$ can be 
found from the exact result given in Appendix A. It is equal to $c = 1/64\pi$.

Similarly, 
the amplitude ${\cal M}_{\phi N \chi N}$ is, in the large $M_{N}$ limit,
 equal to
${\cal M}_{\phi N H N}$.
Dimensional analysis can also be applied to the remaining boxes, confirming that
${\cal M}_{\phi N H N}$ and ${\cal M}_{\phi N \chi N}$ are the only box
diagrams with quadratic nondecoupling.
%**************************************

\subsection{Vertex diagrams}

We next consider together vertex corrections and corrections to the external
charged leptons.
Diagrams modifying the $W\mu\nu_{i}$ vertex are depicted in Fig.~\ref{vermuon}.
 Another set, one that modifies the $We\nu_{j}$
vertex, is not shown.

%\begin{figure}[hbtp]
%\begin{center}
%\setlength{\unitlength}{1in}
%\begin{picture}(6,3)
%\put(0.93,+0.325){\mbox{\epsfxsize=4.0in\epsffile{p2fig2.eps}}}
%\end{picture}
%\end{center}
%\caption{Vertex diagrams for muon decay}
%\label{vermuon}
%\end{figure}
The sum over the depicted set of diagrams gives the muon
vertex amplitude ${\cal M}_{vertex}^{\mu}$:
%%%%%%%%%%%%%%%%%%%%%%%% Vertex  graphs %%%%%%%%%%%%%%%%%%%%%%%%%%%%%%%%%%%%%
\begin{eqnarray}
{\cal M}_{vertex}^{\mu} & = & {\cal M}_{\mu \nu Z} + {\cal M}_{\mu N Z} +
{\cal M}_{Z W \mu} + {\cal M}_{\gamma W \mu} + {\cal M}_{WZ\nu}\nonumber   \\
& + &
{\cal M}_{WZN} + {\cal M}_{\phi ZN} + {\cal M}_{WHN}
+ {\cal M}_{\phi HN} + {\cal M}_{\phi \chi N} \\
& = & 
\Lambda^{\mu} {\cal M}_{tree}. \nonumber  
\end{eqnarray}
Explicit expressions for each of these amplitudes are given in Appendix A.
They are divergent and we renormalize them with the SM form counterterms
\cite{key6} (renormalized quantities are distinguished by the hat):
\begin{eqnarray}
\label{eq9}
{\hat \Lambda}^{\mu} & = & \Lambda^{\mu} + \delta Z_{1}^{W} - \delta Z_{2}^{W}
+ \delta Z_{L}^{\mu},
\end{eqnarray}
where
\begin{eqnarray}
\delta Z_{1}^{W} - \delta Z_{2}^{W} & = & - \frac{\alpha}{2 \pi s_{W}^{2}} (\frac{2}{\epsilon} -\gamma + \ln 4\pi - \ln \frac{M_{W}^{2}}{\mu^{2}}) \; = \; - \frac{\alpha}{2 \pi s_{W}^{2}} \Delta_{M_{W}},\\
\label{finans}
\delta Z_{L}^{\mu} & = & - \Sigma_{L}^{\mu} (m_{\mu}^{2}) + 
\frac{\alpha}{2 \pi} \Big( 2 \ln \frac{m_{\mu}}{\lambda} - 1 \Big),
\end{eqnarray}
$2/\epsilon$ with $\epsilon \rightarrow 0$ is the pole of the dimensionally regularized amplitudes and $\lambda$ is the regularized photon mass.
$\Sigma_{L}^{\mu} = \Sigma_{L}^{WN} + \Sigma_{L}^{W\nu} +
\Sigma_{L}^{\phi N} + \Sigma_{L}^{Z\mu} + \Sigma_{L}^{\gamma \mu}$ 
is the left-handed part of the muon self-energy, with the individual terms corresponding to the loops shown in Fig.~\ref{selffd}. All these contributions are given in \cite{melo} and \cite{key6}. The term which we use specifically
below, $\Sigma_{L}^{\phi N}$, is given in Appendix A.
 In our scheme, the renormalized
charged lepton self energies do not contribute directly, but rather through
the renormalization constant $\delta Z_{L}$. Cancellation of divergences occurs
as usual between the vertex loops and the counterterm contributions but we
focus here on the $M_N$ dependent terms only.
Looking for the dominant graphs in the limit $M_{N} \gg M_{W},
M_{Z}, M_{H}$, we find (either by taking the limit of exact results or using
dimensional analysis and power counting)
that the graphs of Fig.~\ref{vermuon}f have quadratic nondecoupling.
However, both infinite and 
finite parts of these two graphs are cancelled in the large $M_{N}$ limit
by the $\Sigma_{L}^{\phi N}$ 
term (see Fig.~\ref{selffd}b) in the counterterm $\delta Z_{L}^{\mu}$. Therefore there remain 
no $M_{N}^{2}$ dependent terms in the renormalized vertex diagrams.

%\begin{figure}
%\begin{center}
%\setlength{\unitlength}{1in}
%\begin{picture}(6,2)
%\put(.38,+0.325){\mbox{\epsfxsize=5.0in\epsffile{p2fig3.eps}}}
%%\put(1.65,0){\footnotesize {\bf Figure \ref{selffd}:} Lepton self-energies.}
%\end{picture}
%\end{center}
%\caption{Charged lepton self-energies.}
%\label{selffd}
%\end{figure}
            
This curious cancellation can be seen either explicitly (both $\Sigma_{L}^{\phi
N}$ and ${\cal M}_{\phi HN}$, ${\cal M}_{\phi \chi N}$ are given in Appendix A)
or, better yet, after applying the symmetries of the theory.  The way to go is
to study the more familiar case of a $\gamma l l$ vertex.
This vertex is modified from its tree-level value $i e \gamma_{\mu}$ by the
one-loop diagrams shown in Fig.~\ref {gll} as (we show only vector and axial
vector corrections)
\begin{eqnarray}
i e \gamma_{\mu} & \rightarrow &  i e \gamma_{\mu}(1 + F_{V}) - i e
\gamma_{\mu}\gamma_{5} F_{A}.
\end{eqnarray}
%\begin{figure}
%\begin{center}
%\setlength{\unitlength}{1in}
%\begin{picture}(6,3)
%\put(.38,+0.325){\mbox{\epsfxsize=5.0in\epsffile{p2fig4.eps}}}
%%\put(1.65,0){\footnotesize {\bf Figure \ref{selffd}:} Lepton self-energies.}
%\end{picture}
%\end{center}
%\caption{$\gamma l l$ vertex .}
%\label{gll}
%\end{figure}
We now use a Ward-identity \cite{key6,wst}, which
relates the vertex formfactors $F_{V,A}$ evaluated at
$(p_{1}+p_{2})^{2}=0$ ($p_{1}, p_{2}$ are lepton momenta)
to charged lepton self-energies
represented by the counterterms $\delta Z_{V,A}^{l}$:
\begin{eqnarray}
\label{wardi}
F_{V,A}(0) + \delta Z_{V,A}^{l} & = & \frac{1}{4s_{W}c_{W}}
\frac{\Sigma_{\gamma Z}(0)}{M_{Z}^{2}},
\end{eqnarray}
where 
$\delta Z_{V}^{l} = \frac{1}{2}(\delta Z_{L}^{l} + \delta Z_{R}^{l})$,
$\delta Z_{A}^{l} = \frac{1}{2}(\delta Z_{L}^{l} - \delta Z_{R}^{l})$,
and $\Sigma_{\gamma Z}(0) = \frac{\alpha}{2\pi}\frac{M_{W}^{2}}{c_{W}s_{W}}\Delta_{M_{W}}$ is the term originating in the bosonic loops of
the $\gamma$-Z mixing.
At small $M_{N}$ the graphs with unphysical Higgs $\phi$ are negligible,
however, with $M_{N} \gg M_{W}, M_{Z}, M_{H}$ two types of graphs dominate the left-hand side of 
Eq.~\ref{wardi}: the irreducible vertex (formfactor) $F_{V,A}^{\phi \phi N}$ 
(see Fig.~\ref{gll}c) 
and the self-energy (its vector or axial-vector part) $\delta Z_{V,A}^{\phi N}$
(Fig.~\ref{selffd}b). 
Since the right-hand side of Eq.~\ref{wardi} is not
affected by the NHL's, it remains constant and (very) small with respect to
$F_{V,A}^{\phi \phi N}$ or $\delta Z_{V,A}^{\phi N}$
at $M_{N} = O$(TeV).
Hence the only way to meet the
above formula is to have
\begin{eqnarray} 
\label{hophop}
F_{V,A}^{\phi \phi N} + \delta Z_{V,A}^{\phi N} & = & 0
\end{eqnarray}
%${\cal M}_{\phi N}^{\gamma} = - {\cal M}_{\phi \phi
%N}^{\gamma}$
in the limit of large $M_{N}$.
If we now return from the $\gamma l l$ vertex to the $W \mu \nu$ vertex, 
we find a similar result (for proof see Appendix B):
\begin{eqnarray}
\label{for1}
\Lambda_{\phi H N} + \Lambda_{\phi \chi N} + \delta Z_{L}^{\phi N} & = & 0,
\end{eqnarray}
that is, the two dominant nonstandard contributions from Eq.~\ref{eq9} cancel exactly, including the finite parts. Since the remaining nonstandard contributions are not enhanced by the quadratic nondecoupling and are suppressed by the mixings, vertices can be reliably represented by the SM terms.

\subsection{Neutrino self-energy and its renormalization}
\label{secneus}

Half the neutrino self-energy diagrams contributing to muon decay
are shown in Fig.~\ref{selfmuon}. 
The corresponding self-energy is denoted as $\Sigma^{\nu_{\mu}}$. 
The other half consists
of the same loops sitting on the bottom neutrino leg with the corresponding
self-energy $\Sigma^{\nu_{e}}$.
In all these diagrams, we sum over the internal massless neutrinos 
$\nu_{k}, k=1,2,3$. In principle,
the graphs with $\nu_{k}$ replaced by $N_{a}$ are also present, however, they
are suppressed by the large mass $M_{N}$.

%\begin{figure}[hbtp]
%\begin{center}
%\setlength{\unitlength}{1in}
%\begin{picture}(6,2)
%\put(.4,+0.325){\mbox{\epsfxsize=5.0in\epsffile{p2fig5.eps}}}
%\end{picture}
%\end{center}
%\caption{Neutrino self-energy diagrams for muon decay}
%\label{selfmuon}
%\end{figure}

%%%%%%%%%%%%%%%%%%% Neutrino selfenergy %%%%%%%%%%%%%%%%%%%%%%%%%%%%%%%%%%
The unrenormalized
neutrino self-energy $\Sigma^{\nu_{l}}$ ($l = e, \mu$) has the form
\begin{equation}
\Sigma^{\nu_{l}}  =  \frac{1}{2}\Sigma_{L}^{\nu_{l}}
\not p (1-\gamma_{5}),
\end{equation}
where $\Sigma_{L}^{\nu_{l}}$ receives contributions (given in Appendix A) 
from the diagrams of Fig.~\ref{selfmuon}.
The amplitude for those diagrams, in terms of
$\Sigma_{L}^{\nu_{l}}$, can be shown to be equal to
\begin{equation}
{\cal M}_{self} = - {\cal M}_{tree}\frac{\Sigma_{L}^{\nu_{l}}}{2},
\end{equation}
where the factor $\frac{1}{2}$ comes from our dealing with the external wave
function rather than the neutrino propagator.

Let us now investigate the question of the renormalization of $\Sigma^{\nu_{l}}$. In this case the counterterms
are modified from their SM form. 
%\footnote{So far we have used
%the SM form of the counterterms, see Appendix \ref{recons}. The actual value 
%of the counterterms was, of course, different from SM.}.
The problem is how to
renormalize a part of a theory where interaction eigenstates are different 
from mass eigenstates. Curiously, this also happens in the SM quark sector \cite{sack}. The
difference is that in the SM the problem is circumvented by arguing the 
off-diagonal quark mixings are too small  
to have any effect in the loops and the
renormalization procedure is effectively simplified to that of mass eigenstates
being also flavour eigenstates.  In our model, we cannot neglect the 
`off-diagonal'
mixings (their role is assumed by $ll_{mix}$), since they (in combination with TeV NHL masses)
lead to the dominant terms in the predicted deviation from SM results.
This problem was studied in Refs.~\cite{kniehl,sack}. 
In Ref.~\cite{kniehl} it was shown that in general the renormalization of the
divergent amplitudes requires the renormalization of the mixing matrix. In our model, the amplitudes can be renormalized without the renormalization of the mixing matrix, if the assumption
of zero flavour-violating mixing parameters is made. Our scheme is a straightforward extension of the SM counterterm.

We start with the counterterm Lagrangian,
which has the same form as that of the SM.
\begin{eqnarray}
\label{ourcounter}
i \;\delta Z_{L}^{e} \; \overline{\nu}_{e}  {\not \partial} \nu_{e}
+ i\; \delta Z_{L}^{\mu} \overline{\nu}_{\mu}  {\not \partial} \nu_{\mu}
+ i \;\delta Z_{L}^{\tau} \overline{\nu}_{\tau}  \not {\partial} \nu_{\tau}.
\end{eqnarray}
Weak eigenstates $\nu_{l}$ are given in terms 
of mass eigenstates $\nu_{i}, N_{a}$ in Eq.~\ref{alter}.
This gives us, for the product $\overline{\nu}_{l}\nu_{l}$,
\begin{eqnarray}
\overline{\nu}_{l}\nu_{l} & = &
\sum_{k,i=1,2,3}\overline{\nu}_{i}\big(K_{L}^{\dagger}\big)_{il}
\big(K_{L}\big)_{lk}\nu_{k} + ... (\overline{\nu}_{i}N, \overline{N}\nu_{k},
\overline{N}N),
\end{eqnarray}
and Eq.~\ref{ourcounter} thus contributes the following massless neutrino counterterm:
\begin{equation}
\sum_{k,i=1,2,3}\Big\{\delta
Z_{L}^{e}\big(K_{L}^{\dagger}\big)_{ie}\big(K_{L}\big)_{ek}
+ \delta Z_{L}^{\mu}\big(K_{L}^{\dagger}\big)_{i\mu}\big(K_{L}\big)_{\mu k}
+ \delta Z_{L}^{\tau}\big(K_{L}^{\dagger}\big)_{i\tau}\big(K_{L}\big)_{\tau k}
\Big\}\overline{\nu}_{i} {\not \partial} \nu_{k}.
\end{equation}
In our case we sum over internal $\nu_{k}$ but not over external
$\nu_{i}$.
The graphic representation of the relevant counterterm (embedded in muon decay)
is in Fig.~\ref{countl}.

%\begin{figure}[hbtp]
%\begin{center}
%\setlength{\unitlength}{1in}
%\begin{picture}(6,2)
%\put(1.4,+0.025){\mbox{\epsfxsize=2.8in\epsffile{p2fig6.eps}}}
%\end{picture}
%\end{center}
%\caption{Counterterm diagram for neutrino self-energy in muon decay}
%\label{countl}
%\end{figure}

The amplitude for this diagram is
%Including the factor $\big(K_{L}^{\dagger}\big)_{i\mu}$, the new counterterm
%Lagrangian for the massless neutrinos is  given as
\begin{equation}
{\cal M}_{C} = - \frac{1}{2}{\cal M}_{tree}^{SM}\sum_{l=e,\mu,\tau}
\sum_{k=1,2,3} \delta Z_{L}^{l}
\big(K_{L}^{\dagger}\big)_{il}\big(K_{L}\big)_{lk}
\big(K_{L}^{\dagger}\big)_{k\mu}\big(K_{L}\big)_{ej}.
\end{equation}
Again, the factor $\frac{1}{2}$
comes from our dealing with the external wave function
rather than the internal propagator.
${\cal M}_{tree}^{SM}$ is the tree level amplitude for muon decay in the SM.
The mixing factors 
$\big(K_{L}^{\dagger}\big)_{k\mu}$ and $\big(K_{L}\big)_{ej}$ 
originate at the $\mu W \nu_{k}$ and $e W \nu_{j}$
vertices, respectively.  The amplitude ${\cal M}_{C}$ can be further simplified,
\begin{eqnarray}
{\cal M}_{C} & = & - \frac{1}{2}{\cal M}_{tree}^{SM} \sum_{l=e,\mu,\tau}
\delta Z_{L}^{l}\big(K_{L}^{\dagger}\big)_{il}
\sum_{k=1,2,3}\big(K_{L}\big)_{lk}\big(K_{L}^{\dagger}\big)_{k\mu}
\big(K_{L}\big)_{ej} \nonumber \\
& = & - \frac{1}{2}
{\cal M}_{tree}^{SM} \sum_{l=e,\mu,\tau}\delta
Z_{L}^{l}\big(K_{L}^{\dagger}\big)_{il} \big(\delta_{l\mu} - l\mu_{mix}\big)
\big(K_{L}\big)_{ej} \nonumber \\ 
& = & -  \frac{1}{2}
{\cal M}_{tree}^{SM} \delta Z_{L}^{\mu} \big(1 - \mu \mu_{mix} \big)
\big(K_{L}^{\dagger}\big)_{i\mu}\big(K_{L}\big)_{ej} \nonumber \\
& = & - \frac{1}{2}
\delta Z_{L}^{\mu} \big(1 - \mu \mu_{mix} \big) {\cal M}_{tree}.
\end{eqnarray}
The factor $\big(K_{L}^{\dagger}\big)_{i\mu}\big(K_{L}\big)_{ej}$
 was absorbed by ${\cal M}_{tree} =
{\cal M}_{tree}^{SM} \big(K_{L}^{\dagger}\big)_{i\mu}\big(K_{L}\big)_{ej}$.

Now we can write down the final expressions for the renormalized amplitude 
${\hat {\cal M}}_{self}$  and the renormalized neutrino self-energy ${\hat \Sigma_{L}}^{\nu_{l}}$ :  
\begin{eqnarray}
\label{modren}
{\hat {\cal M}}_{self} & = & {\cal M}_{self} + {\cal M}_{C} \; = \;
- \frac{\Sigma_{L}^{\nu_{l}}}{2} {\cal M}_{tree} - \frac{\delta Z_{L}^{l}}{2}
\big(1 - ll_{mix}\big) {\cal M}_{tree}, \\
\label{modren1}
{\hat \Sigma_{L}}^{\nu_{l}} & = & \Sigma_{L}^{\nu_{l}} + \delta Z_{L}^{l}
 \big(1 - ll_{mix}\big).
\end{eqnarray}
The constant $\delta Z_{L}^{l}$ was given in Eq.~\ref{finans}.

To prove the cancellation of the infinities, we note that the infinite part of
$\delta Z_{L}^{l}$ is given by \cite{melo,key6}
\begin{eqnarray}
\delta Z_{L}^{l,\infty} & = & - \frac{\alpha}{4 \pi}\frac{1}{s_{W}^{2}}
 \Big\{\frac{1}{2} + \frac{1}{4c_{W}^{2}} + \frac{M_{N}^{2}}{4M_{W}^{2}}ll_{mix}\Big\}
\Delta_{\mu},
\end{eqnarray}
where $\Delta_{\mu} = \frac{2}{\epsilon}-\gamma + \ln 4\pi + \ln \mu^{2}$.
The infinite part of the neutrino self-energy is (see Appendix~A)
\begin{eqnarray}
\Sigma_{L}^{\nu_{l},\infty} & = & \frac{\alpha}{4 \pi}\frac{1}{s_{W}^{2}}
\Big\{\frac{M_{N}^{2}}{4M_{W}^{2}}ll_{mix}(1-ll_{mix}) + \frac{1}{2}(1-ll_{mix})
+ \frac{1}{4c_{W}^{2}}ll_{mix}(1-ll_{mix}) \Big. \nonumber \\
& + &
\Big. \frac{1}{4c_{W}^{2}}(1-ll_{mix})^{2} \Big\} \Delta_{\mu}.
\end{eqnarray}
From the formulae above it can be easily seen that infinities cancel out in
Eq.~\ref{modren1}.

We now investigate the large $M_{N}$ behaviour of the renormalized neutrino
self-energy ${\hat \Sigma}_{L}^{\nu_{l}}$, this time using exact results. The
two diagrams of Fig.~\ref{selfmuon}c contribute to the self energy with an
overall factor of $M_{N}^{2}$.  For large $M_{N}$, the coefficients of these
diagrams contain the functions  \begin{eqnarray}
B_{0}(p;M_{H,Z,W},M_{N}) & \sim &  1 - 2 \ln M_{N}, \nonumber \\
B_{1}(p;M_{H,Z,W},M_{N}) & \sim &  - 0.25 + \ln M_{N}.
\end{eqnarray}
This implies quadratic nondecoupling for
$\Sigma_{L}^{H}(p)$ and $\Sigma_{L}^{\chi}(p)$ such that
\begin{eqnarray}
\Sigma_{L}^{H}(p) + \Sigma_{L}^{\chi}(p) & = & \frac{\alpha}{2\pi}
\frac{1}{4s_{W}^{2}} ll_{mix}(1-ll_{mix}) \frac{M_{N}^{2}}{M_{W}^{2}}
\biggl[ \frac{1}{2}\Delta_{\mu} + \frac{3}{4} - \ln M_{N} \biggr] \;\;\;
\end{eqnarray}
$\Sigma_{L}^{\phi N}$ (see Fig.~\ref{selffd}b), 
which contributes to ${\hat \Sigma}_{L}^{\nu_{l}}$ via the
counterterm $\delta Z_{L}^{l}$ (see Eqs.~\ref{finans},~\ref{modren1}),
is given in Appendix A (Eq.~\ref{slfn}). 
From here we can see that (once again) $\Sigma_{L}^{\phi N}$ not only cancels 
infinities in $\Sigma_{L}^{H}(p)$ and $\Sigma_{L}^{\chi}(p)$, but,
in the large $M_{N}$ limit investigated, it also cancels the finite parts.
As a result, there is no quadratic nondecoupling in the renormalized
neutrino self-energy and, as in the case of irreducible vertex corrections,
it suffices to consider just the SM loops.

%************    Results    **********************
\subsection{Results}

The loop corrections to muon decay modify the quantity $\Delta r$ in the
implicit relation between $M_{W}$ and $G_{\mu}$ as follows \cite{melo}
\begin{eqnarray}
\label{mwg}
M_{W}^{2} s_{W}^{2} & = & \frac{\pi\alpha}{\sqrt{2} G_{\mu} (1-\Delta r)}
\Big(1-\frac{1}{2}ee_{mix}-\frac{1}{2}\mu\mu_{mix}\Big),
\end{eqnarray}
where $1-\frac{1}{2}ee_{mix}-\frac{1}{2}\mu\mu_{mix}$ is the tree-level
correction in our model
and $\Delta r$ can be written as 
\begin{eqnarray}
\Delta r & = & \frac{{\cal R}e \:{\hat \Sigma}_{W}(0)}{M_{W}^{2}} + \delta_{V}.
\end{eqnarray}
${\hat \Sigma}_{W}(0)$ is the renormalized self-energy of the $W$ boson
 which we previously calculated \cite{melo}.
The parameter $\delta_{V}$ is the sum of the boxes, irreducible vertices and
self-energies calculated in the previous sections, along with the equivalent
contributions to the $W e \nu$ vertex, 
\begin{eqnarray}
\delta_{V} & = & \frac{{\cal M}_{\gamma e W \mu}+{\cal M}_{box}}{{\cal M}_{tree}}
+ {\hat \Lambda}^{\mu} + {\hat \Lambda}^{e} - 
\frac{1}{2} {\hat \Sigma}^{\nu_{e}}
- \frac{1}{2} {\hat \Sigma}^{\nu_{\mu}}.
\end{eqnarray}
Based on the previous sections, we expect that $\delta_{V}$ can be reliably represented as
\begin{eqnarray}
\label{rel1}
\delta_{V} & \doteq & \delta_{V}^{SM} + \delta_{b}^{e\mu} \; = \; 
\delta_{V}^{SM} + \frac{\alpha}{64 \pi
s_{W}^{2}}\frac{M_{N}^{2}}{M_{W}^{2}} ee_{mix} \mu\mu_{mix},
\end{eqnarray}
where $\delta_{V}^{SM}$ is the SM value \cite{key6} and the rest comes from
just two box diagrams (Fig.~\ref{boxmuon}e).

Numerical results for the corrections to muon decay are shown in Table~\ref{muonloops}. As input data we used
the following set (henceforth the standard set): $M_{Z} = 91.1884$ GeV, $\alpha ^{-1} = 137.036$,
$A \equiv \frac{\pi \alpha}{\sqrt{2} G_{\mu}} = 37.281 \: {\rm GeV}$,
$M_{H} =  200$ GeV, $m_{t} = 176$ GeV. The mixing parameters used are $ee_{mix} = 0.0071$ and $\mu\mu_{mix} = 0.0014$ while for $\tau\tau_{mix}$ we show results for both the minimal and the maximal value allowed,
0 and 0.033, respectively. 
%We suppress $\tau\tau_{mix}$, which at
%its maximal value currently allowed would make corrections 
%to ${\hat \Sigma}_{W}(0)/M_{W}^{2}$ much larger than the
%corrections  to $\delta_{V}$
%(as these latter only depend on $ee_{mix},\mu\mu_{mix}$) and we would like to
%show the case when also the latter are important.
$\tau\tau_{mix}$ is the least well constrained mixing however there is no particular theoretical motivation to assume that it is actually larger than the other mixings. Hence we give results with $\tau\tau_{mix}$ suppressed in order to illustrate the dependence on $ee_{mix}$ and $\mu\mu_{mix}$.
For $\tau\tau_{mix} = 0$ 
in the first three lines of the table we show the contributions of the
self-energy, vertex and box diagrams to $\delta_{V}$ (line 4) for
NHL masses $M_{N}$ of up to 30 TeV.
Also shown (lines 5,6) are ${\hat \Sigma}_{W}(0)/M_{W}^{2}$ and $\Delta r$
 since, 
ultimately, we are interested in NHL effects in the observable $M_{W}$ (line 7).
 The SM values are given in the first column.
The results confirm expectations from the previous sections. There is no
nondecoupling for self-energies and vertices and there is a quadratic
dependence on $M_{N}$, in the large $M_{N}$ limit, for the boxes. The boxes are
becoming important at very high masses. Still, they are small compared to the
change in ${\hat \Sigma}_{W}(0)/M_{W}^{2}$. This is due to the fact that the
dominant boxes enter with the coefficient $ee_{mix}\mu\mu_{mix}$ (see Eq.~\ref{pnhn}), while the correction to the $W$ propagator is proportional to 
$k_{HH} = ee_{mix}^{2} + \mu\mu_{mix}^{2} + \tau\tau_{mix}^{2}$ \cite{melo}, 
which is allowed to be larger given the current bounds on the mixings.
The $W$ mass jumps from $M_{W}^{SM} = 80.459$ GeV to $M_{W} = 80.537$ GeV
at $M_{N} = 0.5$ TeV mainly as a result of the tree-level correction factor
$\Big(1-\frac{1}{2}ee_{mix}-\frac{1}{2}\mu\mu_{mix}\Big)$ (see Eq.~\ref{mwg}).
After that it rises very slowly until the $M_{N}$ dependent
amplitudes become dominant above $5$ TeV.

The results for $\tau\tau_{mix} = 0.033$ case (Table~\ref{muonloops}, lines 8-11) are similar. We only show $\delta_{V}, {\hat \Sigma}_{W}(0)/M_{W}^{2}, \Delta r$ and $M_{W}$ since the boxes, self-energies and vertices change slowly with $\tau\tau_{mix}$ (they only depend implicitly on  $\tau\tau_{mix}$, via $s_{W}$), as illustrated by $\delta_{V}$ in line 8. The relative impact of nonSM boxes ($\delta_{V}$) on  $\Delta r$ compared to that of ${\hat \Sigma}_{W}(0)/M_{W}^{2}$ decreases with increasing $\tau\tau_{mix}$.
${\hat \Sigma}_{W}(0)/M_{W}^{2} \sim \tau\tau_{mix}^{2} M_{N}^{2}$ corrections actually violate (for $M_{N} > 5$~TeV) the perturbative unitarity bound discussed in Ref. \cite{melo} . 

To sum it up,
the analysis of Ref.~\cite{melo} turns out to be basically valid even
after the restriction $ee_{mix} = \mu\mu_{mix} = 0$ is relaxed. The numerical predictions can be
improved by the inclusion of the tree-level correction
$\Big(1-\frac{1}{2}ee_{mix}-\frac{1}{2}\mu\mu_{mix}\Big)$,
while the largest loop corrections, 
the box diagrams of Fig.~\ref{boxmuon}e, are only marginally important.
Only in the case of the Z decay into $e^{+}e^{-}$, with $ee_{mix}$ and $\tau\tau_{mix}$ now made comparable, does the character of the $M_{N}$ dependence change (see below).

The impact of these results is shown in Figs.~\ref{resultsa} and \ref{resultsb}.
We give the Z leptonic widths as a function of NHL mass for the two cases, $ee_{mix}=0$ and $ee_{mix}=0.0071$, ($\mu\mu_{mix}$ is negligible) in
Figs.~\ref{resultsa} and \ref{resultsb}, respectively. These figures show the widths for three values of $\tau\tau_{mix}: 0.033, 0.02$, and $0.01$.
The remaining input data come from the standard set.
The dashed lines represent the $1\sigma$ variation about the current experimental results for the individual $Z$
leptonic widths, $\Gamma_{\tau \tau}^{exp} = 83.85 \pm 0.29$ MeV and $\Gamma_{ee}^{exp} = 83.92 \pm 0.17$ MeV \cite{mt1}. The one-loop SM prediction is $\Gamma_{\tau \tau} = \Gamma_{ee} = 84.03$ MeV.

We have discussed the main features of Figs.~\ref{resultsa}a,b ($\Gamma_{ll} \sim \tau\tau_{mix}^{2} M_{N}^{2}$, rising $\Gamma_{ee}$ vs falling $\Gamma_{\tau \tau}$ for $\tau\tau_{mix}$ dominating $ee_{mix}$) previously \cite{melo}.
Here we point out that the main difference between Figs.~\ref{resultsa}a,b and
Figs.~\ref{resultsb}a,b, namely the total upward shift of the widths in the latter, can indeed be traced to the tree-level correction to the $\mu$-decay. Unfortunately, this tree-level correction interferes destructively with the one-loop corrections which drive $\Gamma_{\tau \tau}$ widths down. Also note that as $\tau\tau_{mix}$ and $ee_{mix}$ become comparable, so do $\Gamma_{\tau \tau}$ and $\Gamma_{ee}$, as expected.

Our best constraints at the $2\sigma$ level on NHL mass come from $\Gamma_{\tau\tau}$, shown in Fig.~\ref{resultsa}a \footnote{Constraints from $\Gamma_{ee}$ shown in Fig.~\ref{resultsa}b are just slightly worse at the $2\sigma$ level.}; we get $M_{N} \leq 4.3$ TeV for $\tau\tau_{mix} = 0.033$, $M_{N} \leq 7$ TeV for $\tau\tau_{mix} = 0.02$, and $M_{N} \leq 13$ TeV for $\tau\tau_{mix} = 0.01$.
These constraints can be neatly summarized in the following approximation
\begin{eqnarray}
\label{mng}
M_{N} & < & 4.3 \times \frac{0.033}{\tau \tau_{mix}} \;{\rm TeV}.
\end{eqnarray}
Here, this assumes $\tau\tau_{mix}$ dominates $ee_{mix}$.
%From the universality breaking parameter we get similar result:
%\begin{eqnarray}
%\label{mnubr}
%M_{N} & < & 3.8 \times \frac{0.033}{\tau \tau_{mix}} \;{\rm TeV}.
%\end{eqnarray}

We note that as the value of $\tau\tau_{mix}$ is more tightly constrained, these limits are less restricted than those from perturbative unitarity considerations \cite{melo}:
\begin{eqnarray}
M_{N} & < & 4 \times \sqrt{\frac{0.033}{\tau \tau_{mix}}} \;{\rm TeV}.
\end{eqnarray}

Finally, we note that our computation of the muon decay loops also enables us to find the one-loop
modification of Eq.~\ref{modif} by NHL's:
\begin{eqnarray}
\frac{\Gamma(\tau \rightarrow e \nu \nu) /
\Gamma^{SM}(\tau \rightarrow e \nu \nu)}
{\Gamma(\mu \rightarrow e \nu \nu) /
\Gamma^{SM}(\mu \rightarrow e \nu \nu)}
& = & \Big(\frac{g_{\tau}}{g_{\mu}}\Big)^{2} \; = \;
\frac{1-\tau\tau_{mix}}{1-\mu\mu_{mix}}\frac{1+2\delta_{b}^{e\tau}}
{1+2\delta_{b}^{e\mu}},
\end{eqnarray}         
where $\delta_{b}$ is given in Eq.~\ref{rel1}.               
For $M_{N} = 4$ TeV, and the current constraints on the mixing parameters, we find that the one-loop correction is only about $1 \%$
of the tree-level correction; therefore the constraints of Ref.~\cite{Nardi}
are indeed independent of the NHL mass.

%Setting $\mu\mu_{mix}=0$, we get
%\begin{eqnarray}
%\tau\tau_{mix} & = & 1 - \Big(\frac{g_{\tau}}{g_{\mu}}\Big)^{2},
%\end{eqnarray}
%with \cite{Nardi}
%\begin{eqnarray}
%\Big(\frac{g_{\tau}}{g_{\mu}}\Big)^{2} & = & 0.989 \pm 0.016 .
%\end{eqnarray}

\section{Flavour-conserving vs flavour-violating decays}
\label{fvzb1}

In this section, we review the constraints on the parameters in our
model as derived from flavour-violating and flavour-conserving decays (the latter will be represented by the leptonic decays of the $Z$ boson).
We compare the sensitivity of these two classes of processes to the presence of NHL's. 

Lepton
flavour-violating decays have so far received a lot more attention
\cite{bernabeu1,Ng1,Ng2,ggjv,Ilakovac,Jarlskog,Valle2,Korner,pilaftsis1}
than the flavour-conserving processes \cite{melo,bernabeu2}. The calculation 
of the flavour-violating processes
is simpler, with a smaller number of contributing diagrams and without
the need to renormalize.
Also there could be a certain preconception that the experimental
signature of the flavour violation is more `dramatic'. It is our intention
to show here that in many cases this expectation is
not justified. We give also the results of our calculation of flavour-violating
decays of the $Z$ boson.
A summary of experimental limits, theoretical predictions and the constraints
on the mixings and/or NHL masses implied by flavour-violating decays is given in Table \ref{tabulka}.
We will now address these decays one by one.

 In the case of the flavour-violating mixing parameters, 
the constraint on one of them, $e \mu_{mix}$,
arises from the measured limit of a rare decay 
$\mu \rightarrow e \gamma$.
%which supplies 
%us with a very stringent limit on $|e\mu_{mix}|$. 
 This decay was studied in the context of our model
and of see-saw models 
with enhanced mixings (an example of a see-saw model with enhanced mixings is the model of Ref.~\cite{pilaftsis2})
by several authors \cite{Ng1,Ng2,ggjv,Ilakovac,Jarlskog}. 
The $\mu \rightarrow e \gamma$ branching ratio goes like $|e \mu_{mix}|^{2}$
times a function which is independent of $M_{N}$ for $M_{N} > 500$ GeV.
The current experimental limit on the $\mu \rightarrow e \gamma$ branching ratio, $BR_{exp} \leq 4.9 \times 10^{-11}\;\;$ \cite{pdb},
%\cite{pdb},
% \begin{eqnarray}
%BR(\mu \rightarrow e \gamma) & \leq & 4.9 \times 10^{-11}\;\;\;\;\;90\%\;
%{\rm C.L.},
%\end{eqnarray}
yields a very stringent upper limit on the mixing of
$|e\mu_{mix}| \leq  0.00024$ (see Table \ref{tabulka}). 

One might expect the other flavour-violating
mixing parameters, $e \tau_{mix}$ and $\mu \tau_{mix}$, to be limited by the
corresponding flavour-violating $\tau$ decays. However, experimental limits on
$\tau \rightarrow e \gamma$ and $\tau \rightarrow \mu \gamma$, 
%\cite {pdb},
%\begin{eqnarray}
%BR_{exp}(\tau \rightarrow e \gamma) & < & 1.2 \times 10^{-4}, 
%\;\;\;\;\;90 \%\;\;{\rm C.L.}\;, \nonumber \\
%BR_{exp}(\tau \rightarrow \mu \gamma) & < & 4.2 \times 10^{-6},
%\;\;\;\;\;90 \%\;\;{\rm C.L.}\;, 
%\end{eqnarray}
$BR_{exp} \leq 1.2 \times 10^{-4}, BR_{exp} \leq 4.2 \times 10^{-6}$ respectively \cite{pdb}, are much weaker.
Moreover, the predicted rate $BR_{th} = 7 \times 10^{-7}$  \cite{ggjv} is now out-of-date due to improved constraints on the mixings (see Table \ref{tabulka}).
%The rate for both $e \gamma$ and $\mu \gamma$ modes was predicted to be
%\cite{ggjv} \begin{eqnarray}
%BR_{th}(\tau \rightarrow e \gamma, \mu \gamma) & = & 
%7 \times 10^{-7}, \;\;\;\;{\rm for}\;\;\;\;M_{N} > 500 \;{\rm GeV}.
%\end{eqnarray}
%However, the limits on mixing parameters used in Ref.~\cite{ggjv}
% are now out of date. With the current limits ($\tau\tau_{mix}=0.033$) the pred%icted rate would be
With the current limit ($\tau\tau_{mix}=0.033$) the predicted rate would be
smaller by at least one order of magnitude, implying that the theoretical result
is two orders of magnitude below the experimental upper limit for  
$\mu \gamma$ mode and about three orders for
$e \gamma$ mode. As a result, it is not these flavour-violating processes which
place the strongest limits on the flavour-violating mixing parameters. Rather, 
for $\mu\tau_{mix}$ and $e\tau_{mix}$, 
we have to use indirect limits obtained by combining the global analysis
results for the flavour-conserving mixing parameters with the inequality
Eq.~\ref{ineq}.

Several other flavour-violating processes at very low energies have been
considered. Another well-constrained muon decay mode is $\mu \rightarrow
e^{-}e^{-}e^{+}$ ($BR_{exp} \leq 1.0 \times 10^{-12}$, \cite {pdb}),
%with \cite {pdb}
%\begin{eqnarray}
%\label{mueee4}
%BR_{exp}(\mu \rightarrow e^{-}e^{-}e^{+}) & < & 1.0 \times
%10^{-12},\;\;\;\;\;90 \% {\rm C.L.}\;.
%\end{eqnarray}
%This process was 
studied in Refs.~\cite{Ng1,Ng2,Ilakovac,Jarlskog}.
The calculation shows the quadratic nondecoupling which we will encounter in
the lepton flavour-violating decays of the $Z$ boson below. 
Ref. \cite{Jarlskog} gives (with an assumption
discussed therein) the following constraint on NHL mass as a function of
$ee_{mix}, e\mu_{mix}$ (see Table \ref{tabulka}):
\begin{eqnarray}
\label{mnlimit4}
M_{N}^{2} & \leq & 0.93 \times 10^{-5} \frac{1 {\rm
TeV}^{2}}{ee_{mix}|e\mu_{mix}|}.
\end{eqnarray}
%which for $M_{N} \geq 3$ TeV is competitive with a constraint implied by Eqs.
%\ref{limits1} and \ref{ineq}:
%\begin{eqnarray}
%ee_{mix}|e\mu_{mix}| & \leq & 0.17 \times 10^{-5}.
%\end{eqnarray}
Also considered in Refs.~\cite{Ng1,Ng2,Jarlskog} is 
$\mu - e$ conversion in nuclei,
$\mu^{-}(A,Z) \rightarrow e^{-}(A,Z)$. The resulting constraint on the product
$ee_{mix}|e\mu_{mix}|$ \cite{Jarlskog} is similar to the one above.

For the flavour-violating decays of the tau into three leptons ($\tau
\rightarrow$ $e^{-}e^{-}e^{+},$ $e^{-}\mu^{-}\mu^{+},$ etc) we know of no
calculation studying the large (TeV) NHL mass limit  in the context of our
model. However, within the see-saw model of Ref.~\cite{pilaftsis2}, Pilaftsis
 \cite{pilaftsis1} predicts with the current limits on mixings ($\tau\tau_{mix}=0.033$) and for $M_{N} = 3$ TeV rates 
$BR_{th}(\tau \rightarrow e^{-}e^{-}e^{+})  =  5 \times 10^{-7}, 
BR_{th}(\tau \rightarrow e^{-}\mu^{-}\mu^{+})  =  3 \times 10^{-7}$ which are well below the current experimental limit $BR_{exp} \leq 1.4 \times 10^{-5}$ \cite{pdb} (see Table \ref{tabulka}).  
%\begin{eqnarray}
%BR_{th}(\tau \rightarrow e^{-}e^{-}e^{+}) & = & 5 \times 10^{-7}, \nonumber   \\
%BR_{th}(\tau \rightarrow e^{-}\mu^{-}\mu^{+}) & = & 3 \times 10^{-7}. 
%\end{eqnarray}
%The current experimental limits are \cite{pdb}
%\begin{eqnarray}
%BR_{exp}(\tau \rightarrow e^{-}e^{-}e^{+}) & < & 1.4 \times 10^{-5},\;\;\;\;\;90
%\%\;\;{\rm C.L.}\;,\nonumber \\
%BR_{exp}(\tau \rightarrow e^{-}\mu^{-}\mu^{+})
%& < & 1.4 \times 10^{-5},\;\;\;\;\;90 \%\;\;{\rm C.L.}\;.
%\end{eqnarray}

Finally, hadronic decay modes of the $\tau$ lepton, $\tau \rightarrow l \eta,
l\pi^{0}$ \cite{ggjv} are also disfavoured by loose limits, e.g. $BR_{exp}(\tau
\rightarrow \mu^{-} \pi^{0}) \leq 4.4 \times 10^{-5}$ \cite{pdb}.

Consider now the case of the flavour-violating leptonic decays of the $Z$ boson.
These rare processes were studied in the context of our model previously 
\cite{bernabeu1,Valle2};
however, the limit of large NHL mass was not fully investigated. This point was
noted in Ref.~\cite{Korner}, where the branching ratios for 
$Z \rightarrow l_{1}^{-} l_{2}^{+}\;
(e^{\pm}\mu^{\mp}, \mu^{\pm} \tau^{\mp}, e^{\pm} \tau^{\mp})$
were derived in the see-saw model of Ref.~\cite{pilaftsis2}.
We therefore here present the results
in our model, having carefully treated the case of a large NHL mass but
without showing the calculational details.  The loop diagrams involved are very
similar to those of the flavour-conserving leptonic decays of the $Z$ boson which
we discussed fully in Ref.~\cite{melo} and the calculation of the
flavour-violating process is very similar to that of Ref.~\cite{Korner} for the
other model. The particular predictions depend, along with the NHL mass, on
the various mixings and their relative phases. We do not present full details
here since the results are not particularly promising. The branching ratio
$BR(l_{1}^{\pm} l_{2}^{\mp}) \equiv \Gamma_{l_{1}^{+} l_{2}^{-} + l_{1}^{-}
l_{2}^{+}}/\Gamma_{Z}$ is shown as a function of $M_{N}$ in Fig.~\ref{fviolfig}a,b. The
parameter $\delta$ introduced in these figures corresponds to the allowed
range of relative phases arising in the mixing factors. In Fig.~\ref{fviolfig}a, we set
$\delta = -1$ and in Fig.~\ref{fviolfig}b, $\delta = +1$.   

As an example for comparison with experiment, using maximally allowed
mixings ($\tau\tau_{mix}=0.033$) and $\delta = +1$, we predict the following branching ratio limits for
$M_{N} = 5$~TeV \begin{eqnarray}
BR_{th}(Z \rightarrow e^{\pm}\mu^{\mp}) & < & 3.3 \times 10^{-8},  \nonumber \\
BR_{th}(Z \rightarrow e^{\pm}\tau^{\mp}) & < & 1.4 \times 10^{-6},   \\
BR_{th}(Z \rightarrow \mu^{\pm}\tau^{\mp}) & < & 2.2 \times 10^{-7}.\nonumber 
\end{eqnarray}
These results are similar to those of 
Ref.~\cite{pilaftsis1}, where, as noted above, the
calculation was done in the context of a see-saw model with enhanced mixings.
For experimental limits see Table \ref{tabulka}.
%Current experimental upper limits are \cite{pdb} 
%\begin{eqnarray}
%\label{brra}
%BR_{exp}(Z \rightarrow e^{\pm}\mu^{\mp}) & < & 6 \times 10^{-6},
%\;\;\;\;\;95 \%\;\;{\rm C.L.},  \nonumber \\
%BR_{exp}(Z \rightarrow e^{\pm}\tau^{\mp}) & < & 1.3 \times 10^{-5},
%\;\;\;\;\;95 \%\;\;{\rm C.L.}, \nonumber \\
%BR_{exp}(Z \rightarrow \mu^{\pm}\tau^{\mp}) & < & 1.9 \times 10^{-5},
%\;\;\;\;\;95 \%\;\;{\rm C.L.}\;.
%\end{eqnarray}
Our most promising prediction, for the $e\tau$ mode, is at least one order
of magnitude below  the experimental limit. Hence the flavour-violating
leptonic decays of the $Z$ boson do not represent a good chance for finding
evidence of NHL's. 
%This is the case since the dominant contribution to
%the total rate for $Z \rightarrow l_{1}^{+}l_{2}^{-} + l_{1}^{-}l_{2}^{+}$,
%depends quartically on small mixings and also quartically on the NHL mass
%$M_{N}$ (the dominant amplitude exhibits quadratic nondecoupling). Further mass
%independent limits on mixings will therefore suppress this dominant contribution
%rather quickly, unless $M_{N}$ is extremely large.

We conclude that the flavour-violating processes give only one (mixing dependent) constraint on NHL mass \footnote{The extremely useful limit on $e\mu_{mix}$ arising from $\mu \rightarrow e \gamma$, is not sensitive to $M_{N}$, for $M_{N} > 500$ GeV.} coming from  $\mu \rightarrow e e e$ (or $\mu - e$ conversion in nuclei), see Eq.~\ref{mnlimit4}. For $ee_{mix}=0.0071$ and $|e\mu_{mix}|=0.00024$, this yields $M_{N}<2.3$ TeV.
For the remaining flavour-violating processes to become sensitive to NHL mass, the experimental 
upper limits would have to be pushed down by at least one order of magnitude for flavour-violating leptonic
decays of the $Z$ boson, and by one to two orders of magnitude for 
flavour-violating decays of the $\tau$ lepton. This most likely requires 
increased high luminosity running at LEP~I energy and a $\tau$ factory
\cite{tfactory}. 

On the other hand, the flavour-conserving processes lead to limits on $M_{N}$ 
summarized in Eq.~\ref{mng}, which for $\tau\tau_{mix}=0.033$ give
$M_{N}<4.3$ TeV.
These limits depend on different mixing parameters than the flavour-violating constraint and thus probe a different part of the mixings vs NHL mass parameter space. A disadvantage of the flavour-violating decays is that they are always proportional to a flavour-violating parameter and this can lead, via the inequality Eq.~\ref{ineq}, to their further suppression with respect to flavour-conserving processes. It is perfectly possible that there might be signatures of the flavour-conserving processes even if there is no sign of the flavour-violating ones.

\section{Conclusions}

In this paper, we have generalized our previous analysis of a model containing
NHL's by relaxing the restriction on mixing parameters $ee_{mix}=\mu\mu_{mix}=0$.
This involved evaluating one-loop corrections to the muon decay which feed into
the input parameter $M_{W}$. We found that two box diagrams exhibit
quadratic nondecoupling but that they are only marginally important numerically.
Hence the numerical results of Ref.~\cite{melo} remain
basically valid, although they can be improved by the inclusion of the tree-level
correction to the muon decay,
$\Big(1-\frac{1}{2}ee_{mix}-\frac{1}{2}\mu\mu_{mix}\Big)$.

The mass $M_{N}$, if larger than $M_{Z}$, can presently 
mainly be probed in radiative
corrections (loops). A traditional approach was mostly limited to hypothetical
lepton flavour-violating processes 
such as $\mu \rightarrow e \gamma,\: \mu, \tau \rightarrow e
e^{+} e^{-}, \: Z \rightarrow e^{\pm}\mu^{\mp}$, etc
\cite{bernabeu1,Ng1,Ng2,ggjv,Ilakovac,Jarlskog,Valle2,Korner,pilaftsis1}. 
We reviewed constraints from these processes in Sec IV.

NHL's
could also induce (again via radiative corrections) deviations from the SM
in currently observed processes, such as those we have previously considered:
 the leptonic widths of the $Z$ boson $\Gamma_{ll}$,
lepton universality breaking parameter $U_{br}$, and the mass of the $W$ boson 
$M_{W}$.                                    
The effect of the NHL mass $M_{N}$ in such radiative corrections is, on the one
hand, 
suppressed by small mixings; on the other hand it is enhanced due to
nondecoupling, the violation of the Appelquist-Carazzone theorem \cite{ac}.
These competing tendencies are reflected by the typical behaviour of the dominant
terms,
\begin{eqnarray}
\label{domterm}
&\sim & (\tau \tau_{mix})^{2} M_{N}^{2}. 
\end{eqnarray}
To make up for the small mixings, only NHL's with masses in the TeV range
can lead to significant deviations from the SM.
In the case of one mixing, $\tau \tau_{mix}$, dominating, we found (see Eq.~\ref{mng}) the following approximate dependence of 
$M_{N}$ on $\tau \tau_{mix}$
($2\sigma$ level):
\begin{eqnarray}
M_{N} & < & 4.3 \times \frac{0.033}{\tau \tau_{mix}} \;{\rm TeV}
\end{eqnarray}
which arises from the consideration of $Z$ leptonic decays.
%\begin{eqnarray}
%M_{N} & < & 3.8 \times \frac{0.033}{\tau \tau_{mix}} \;{\rm TeV}
%\end{eqnarray}
%from the universality breaking parameter. 
We also found some sensitivity of the
$W$ mass to NHL mass and mixings, but these are quite dependent on the top
quark mass so we cannot summarize them in the same way.

These limits on $M_{N}$ are only matched by those
from $\mu \rightarrow e e e$.The flavour-violating decay rates for
$\tau$, which we reviewed in Sec. IV, and for the $Z$ boson, derived in Sec. IV, 
are below the
current experimental sensitivity.
Moreover, the $\mu \rightarrow e e e$ decay
depends only on $ee_{mix}$ and $e\mu_{mix}$, two of the six mixing
parameters, and may be unobservable if $ee_{mix}$ 
and/or $e\mu_{mix}$ are very small. The inequality Eq.~\ref{ineq} can further
suppress the  flavour-violating processes
against the flavour-conserving ones
via the 'conspiracy of the phases' in the sum of complex
terms making up the flavour-violating parameters. 
 
For these reasons, the first signatures of neutral heavy leptons could come
from flavour-conserving observables. At this time, LEP has stopped its runs at
the Z-peak energy 
and is running at $130 - 140$ GeV. It will eventually be producing W
pairs which will allow the mass $M_{W}$ to be measured with a precision of 
$0.044$ GeV \cite{mw2}
(currently $M_{W} = 80.410 \pm 0.180$ \cite{mw1}). 
Combined with more precise measurements of the top quark mass we might be in
a position to place even more stringent limits on NHL masses and mixings from
our prediction of $M_{W}$.

\acknowledgements
This work was funded in part by the Natural Sciences and Engineering Research  
Council of Canada. The authors would
like to thank R.K. Carnegie for many useful conversations. 

\appendix
\section*{A}

%\section{Appendix}

%%%%%%%%%%%%%%%%%%%%% Boxes %%%%%%%%%%%%%%%%%%%%%%%%%%%%%%%%%%%%%%%%%%%%%%%%%
The total contribution of the box diagrams (Figs.~\ref{boxmuon} a-g) is
\begin{eqnarray}
{\cal M}_{box} & = & {\cal M}_{ZeW\mu} +{\cal M}_{W \nu Z\nu} +
{\cal M}_{W\nu Z N} +
{\cal M}_{W N Z\nu}
+ {\cal M}_{W N Z N}  \nonumber \\
& + &
{\cal M}_{\phi N Z N} + {\cal M}_{W N H N} + {\cal M}_{W N \chi N} + 
{\cal M}_{\phi N H N} + {\cal M}_{\phi N \chi N}  \nonumber \\
& + &
{\cal M}_{Z \nu W \mu} + {\cal M}_{Z  N W \mu} + {\cal M}_{W e Z\nu}
+ {\cal M}_{W e Z N} \nonumber   \\
\nonumber \\
& = & {\cal M}_{tree}\frac{\alpha}{4 \pi}\Biggl\{
\frac{-1}{4s_{W}^{2}c_{W}^{2}}M_{W}^{2}\biggl[
4{(-\frac{1}{2}+s_{W}^{2})}^{2}{\cal I}_{0}
 + {\cal I}_{0}(1-{\mu \mu}_{mix}) (1-ee_{mix})
 \Biggr. \biggr.   \nonumber \\
& + & 
{\cal I}_{1}(M_{Z})(1-{\mu \mu}_{mix})ee_{mix}
+{\cal I}_{1}(M_{Z}){\mu \mu}_{mix}(1-ee_{mix})
+{\cal I}_{2}(M_{Z}) ee_{mix} \nonumber \biggr. \\
& \times & 
\biggl. {\mu \mu}_{mix}  \biggr]
+ \frac{1}{4 s_{W}^{2}}M_{N}^{4}\biggl[
\frac{1}{c_{W}^{2}} {\cal I}_{3}(M_{Z}) + {\cal I}_{3}(M_{H}) +
{\cal I}_{3}(M_{Z}) - \frac{1}{4 M_{W}^{2}} {\cal I}_{2}(M_{H})\biggr.
\nonumber \\
& - & 
\frac{1}{4 M_{W}^{2}}{\cal I}_{2}(M_{Z})\biggr]ee_{mix}
{\mu\mu}_{mix} +\frac{2(-\frac{1}{2}+s_{W}^{2})}{s_{W}^{2} c_{W}^{2}}M_{W}^{2}
 \biggl[ {\cal I}_{0}(1-ee_{mix}) \nonumber  \\
& + & 
{\cal I}_{1}(M_{Z})ee_{mix} +
{\cal I}_{0}(1-{\mu \mu}_{mix}) 
+ {\cal I}_{1}(M_{Z}){\mu \mu}_{mix} \biggr] \Biggr\},  
\end{eqnarray}
where the integrals ${\cal I}_{0}, {\cal I}_{1}(m), {\cal I}_{2}(m), 
{\cal I}_{3}(m)$ are
\begin{eqnarray}
{\cal I}_{0} & = & \frac{{(4\pi)}^{2}}{i}\int \frac{d^{4}k}{{(2\pi)}^{4}}
\frac{1}{k^{2}(k^{2}-M_{W}^{2})(k^{2}-M_{Z}^{2})} \; = \;
\frac{1}{M_{Z}^{2}-M_{W}^{2}}\ln
\frac{M_{W}^{2}}{M_{Z}^{2}}, \\
& & \nonumber \\
{\cal I}_{1}(m) & = & \frac{{(4\pi)}^{2}}{i}\int \frac{d^{4}k}{{(2\pi)}^{4}}
\frac{1}{(k^{2}-M_{N}^{2})(k^{2}-M_{W}^{2})(k^{2}-m^{2})} \nonumber \\
& & \nonumber \\
& = &
\frac{1}{m^{2}-M_{W}^{2}}
\biggl\{\ln\frac{M_{W}^{2}}{m^{2}} +
\frac{M_{N}^{2}}{M_{W}^{2}-M_{N}^{2}} \ln \frac{M_{W}^{2}}{M_{N}^{2}} -
\frac{M_{N}^{2}}{m^{2}-M_{N}^{2}} \ln
\frac{m^{2}}{M_{N}^{2}}\biggr\},\;\;\;\;\;
\\
& & \nonumber \\
{\cal I}_{2}(m) & = & \frac{{(4\pi)}^{2}}{i}\int \frac{d^{4}k}{{(2\pi)}^{4}}
\frac{k^{2}}{(k^{2}-M_{N}^{2})^{2}(k^{2}-M_{W}^{2})(k^{2}-m^{2})} \nonumber \\
& & \nonumber \\
& = &
\frac{1}{m^{2}-M_{W}^{2}}
\biggl\{ \frac{1}{1-\frac{M_{W}^{2}}{M_{N}^{2}}} +
\frac{\frac{M_{W}^{4}}{M_{N}^{4}} \ln \frac{M_{W}^{2}}{M_{N}^{2}}}
{{(1-\frac{M_{W}^{2}}{M_{N}^{2}})}^{2}}
-\frac{1}{1-\frac{m^{2}}{M_{N}^{2}}} \nonumber \\
& - &
\frac{\frac{m^{4}}{M_{N}^{4}} \ln \frac{m^{2}}{M_{N}^{2}}}
{{(1-\frac{m^{2}}{M_{N}^{2}})}^{2}} \biggr\}
\\
& & \nonumber \\
{\cal I}_{3}(m) & = & \frac{{(4\pi)}^{2}}{i}\int \frac{d^{4}k}{{(2\pi)}^{4}}
\frac{1}{(k^{2}-M_{N}^{2})^{2}(k^{2}-M_{W}^{2})(k^{2}-m^{2})}
 \nonumber \\
& & \nonumber \\
& = &
\frac{1}{m^{2}-M_{W}^{2}}
\biggl\{ \frac{1}{M_{N}^{2}-M_{W}^{2}} +
\frac{M_{W}^{2} \ln \frac{M_{W}^{2}}{M_{N}^{2}}}{{(M_{N}^{2}-M_{W}^{2})}^{2}}
%\biggr. \nonumber \\
%& & \hspace*{2.in} \biggl.
-\frac{1}{M_{N}^{2}-m^{2}} \nonumber \\
& - &
\frac{m^{2} \ln \frac{m^{2}}{M_{N}^{2}}}{{(M_{N}^{2}-m^{2})}^{2}}
\biggr\}. 
\end{eqnarray}
%For the calculation of these integrals see Appendix \ref{Decko}.

%************** Vertex diagrams **************
The computation of the vertex diagrams (Figs.~\ref{vermuon}a-f) yields
\begin{eqnarray}
{\cal M}_{vertex}^{\mu} & = & {\cal M}_{\mu \nu Z} + {\cal M}_{\mu N Z} +
{\cal M}_{Z W \mu} + {\cal M}_{\gamma W \mu} + {\cal M}_{WZ\nu}\nonumber   \\
& + &
{\cal M}_{WZN} + {\cal M}_{\phi ZN} + {\cal M}_{WHN}
+ {\cal M}_{\phi HN} + {\cal M}_{\phi \chi N} \nonumber   \\
& = & {\cal M}_{tree} \frac{\alpha}{4 \pi} \Biggl\{
\frac{2s_{W}^{2}-1}{4s_{W}^{2}c_{W}^{2}}\Big(\Delta_{M_{Z}}-\frac{1}{2}\Big)
(1-{\mu \mu}_{mix}) \Biggr.\nonumber \\
& + & 
\frac{2s_{W}^{2}-1}{4s_{W}^{2}c_{W}^{2}}\Big(\Delta_{M_{Z}}-\frac{1}{2}-
\frac{M_{N}^{2}}{M_{Z}^{2}-M_{N}^{2}}\ln \frac{M_{Z}^{2}}{M_{N}^{2}}\Big)
{\mu \mu}_{mix}\nonumber     \\
& + & 
\frac{\frac{1}{2}-s_{W}^{2}}{s_{W}^{2}}\Big(3 \Delta_{M_{W}} + \frac{5}{2} +
\frac{3}{s_{W}^{2}}\ln c_{W}^{2}\Big) 
+ 3\Big(\Delta_{M_{W}} + \frac{5}{6}\Big)\nonumber \\
& + & 
\frac{3}{2s_{W}^{2}}\Big(\Delta_{M_{W}} + \frac{5}{6} + \frac{1}{s_{W}^{2}}
\ln c_{W}^{2}\Big)(1-{\mu \mu}_{mix})\nonumber  \\
& + & 
\frac{3}{2s_{W}^{2}}\biggl[ \Delta_{M_{W}} + \frac{5}{6} +
\frac{1}{s_{W}^{2}}\ln c_{W}^{2} + \frac{M_{N}^{2}}{M_{Z}^{2}-M_{W}^{2}}
v(M_{Z})\biggr]{\mu \mu}_{mix} \nonumber \\
& + & 
\frac{1}{2c_{W}^{2}} \frac{-M_{N}^{2}}{M_{Z}^{2}-M_{W}^{2}}v(M_{Z})
{\mu \mu}_{mix}\nonumber  \\
& + & 
\frac{1}{2s_{W}^{2}} \frac{-M_{N}^{2}}{M_{H}^{2}-M_{W}^{2}}v(M_{H})
{\mu \mu}_{mix}\nonumber  \\
& + & 
\frac{1}{8s_{W}^{2}}\frac{M_{N}^{2}}{M_{W}^{2}}\biggl[
\Delta_{M_{W}} + \frac{3}{2} -\frac{M_{H}^{2}}{M_{W}^{2}-M_{H}^{2}}\ln \frac
{M_{W}^{2}}{M_{H}^{2}} + \frac{M_{N}^{2}}{M_{H}^{2}-M_{W}^{2}}v(M_{H})
\biggr]{\mu \mu}_{mix}\nonumber  \\
& + & 
\frac{1}{8s_{W}^{2}}\frac{M_{N}^{2}}{M_{W}^{2}}\biggl[
\Delta_{M_{W}} + \frac{3}{2} -\frac{M_{Z}^{2}}{M_{W}^{2}-M_{Z}^{2}}\ln \frac
{M_{W}^{2}}{M_{Z}^{2}} + \frac{M_{N}^{2}}{M_{Z}^{2}-M_{W}^{2}}v(M_{Z})
\biggr]
\nonumber \\
& \times & {\mu \mu}_{mix}  \Biggr\},  
\end{eqnarray}
where 
\begin{eqnarray}
v(m) & = & \ln \frac{M_{W}^{2}}{m^{2}} +
\frac{M_{N}^{2}}{M_{W}^{2}-M_{N}^{2}}\ln \frac{M_{W}^{2}}{M_{N}^{2}}
- \frac{M_{N}^{2}}{m^{2}-M_{N}^{2}}\ln \frac{m^{2}}{M_{N}^{2}}. 
\end{eqnarray}

The part of the charged lepton self-energy which we specifically use
in the text is:
\begin{eqnarray}
\label{slfn}
\Sigma_{L}^{\phi N} & = &  + \frac{\alpha}{16\pi s_{W}^{2}}ll_{mix} 
\frac{M_{N}^{2}}{M_{W}^{2}}
\biggl[ \Delta_{\mu} + \frac{3}{2}  - 2 \ln M_{N} \biggr].
\end{eqnarray}

%****** neutrino self energy *****************
The left-handed part of the neutrino self-energy (Fig.~\ref{selfmuon})
is given by
\begin{eqnarray}
\label{neuself}
\Sigma_{L}^{\nu_{l}} & = &
\Sigma_{L}^{H}(p)+\Sigma_{L}^{\chi}(p)+\Sigma_{L}^{Z,N}(p)
+\Sigma_{L}^{Z,\nu}(p)+
        \Sigma_{L}^{W}(p) \nonumber \\
& = &
\frac{\alpha}{2\pi}(1-ll_{mix})\Biggl\{
\frac{1}{8s_{W}^{2}}
ll_{mix}
\frac{M_{N}^{2}}{M_{W}^{2}}
\biggl[ \frac{1}{2}\Delta_{\mu}+
B_{0}^{fin}(p;M_{H},M_{N})+B_{1}^{fin}(p;M_{H},M_{N})\biggr]\nonumber \\
& + &
\frac{1}{8s_{W}^{2}}ll_{mix} \frac{M_{N}^{2}}{M_{W}^{2}}
\biggl[ \frac{1}{2}\Delta_{\mu}+
B_{0}^{fin}(p;M_{Z},M_{N})+B_{1}^{fin}(p;M_{Z},M_{N})\biggr]\nonumber \\
& + &
\frac{1}{4s_{W}^{2}c_{W}^{2}}ll_{mix}
\biggl[ \frac{1}{2}\Delta_{\mu}-\frac{1}{2}+
B_{0}^{fin}(p;M_{Z},M_{N})+B_{1}^{fin}(p;M_{Z},M_{N})\biggr]\nonumber \\
& + &
\frac{1}{4s_{W}^{2}c_{W}^{2}}(1-ll_{mix})
\biggl[ \frac{1}{2}\Delta_{\mu}-\frac{1}{2}+
B_{0}^{fin}(p;M_{Z},0)+B_{1}^{fin}(p;M_{Z},0)\biggr] \nonumber \\
& + &
\frac{1}{2s_{W}^{2}}
\biggl[ \frac{1}{2}\Delta_{\mu}-\frac{1}{2}+
B_{0}^{fin}(p;M_{W},m_{l}\rightarrow 0)+B_{1}^{fin}(p;M_{W},m_{l}\rightarrow 0)
\biggr]\Biggr\}.
\end{eqnarray}
Here $s = p^{2} = 0 \;\; \ll \;\; M_{H}^{2},M_{Z}^{2},M_{W}^{2},M_{N}^{2}$.

%*************B functions definition *****************
The functions $B_{0}$ and $B_{1}$ are defined as ($\Delta = \frac{2}{\epsilon}-\gamma-\ln \pi$):
\begin{eqnarray}
\label{abfunc}
B_{0}(p;m_{1},m_{2}) & = & \int \frac{d^{n}q}{i\pi^{2}}
\frac{1}{(q^{2}-m_{1}^{2}+i\epsilon)\big[(q-p)^{2}-m_{2}^{2}+i\epsilon\big]}
\nonumber  
\\
& = & \Delta + B_{0}^{fin}(p;m_{1},m_{2}),
 \nonumber \\
B_{0}^{fin}(p;m_{1},m_{2}) & = & -\int_{0}^{1} dx\; 
\ln \big[p^{2}x^{2} + m_{1}^{2} - (p^{2}+m_{1}^{2}-m_{2}^{2})x\big],
 \nonumber \\   \nonumber \\
B_{\mu}(p;m_{1},m_{2}) & = & \int \frac{d^{n}q}{i\pi^{2}}
\frac{q_{\mu}}{(q^{2}-m_{1}^{2}+i\epsilon)\big[(q-p)^{2}-m_{2}^{2}+i\epsilon
\big]}
\;\;=\;\;-p_{\mu}B_{1}, \nonumber \\
B_{1}(p;m_{1},m_{2}) & = & -\frac{1}{2}\Delta
+ B_{1}^{fin}(p;m_{1},m_{2}), \nonumber \\
B_{1}^{fin}(p;m_{1},m_{2}) & = & \int_{0}^{1} dx \; \ln \big[p^{2}x^{2} + 
m_{1}^{2}-(p^{2}+m_{1}^{2}-m_{2}^{2})x\big] x.   
\end{eqnarray}

%%%%%%%%%%%%%B and F for small s %%%%%%%%%%%%%%%%%%%%%%%%%%%%%%%%%%%%%%%%
For $s = p^{2}$ small with respect to $m_{1}^{2}, m_{2}^{2}, m^{2}$,  we have
\begin{eqnarray}
\label{males}
B_{0}(p;m_{1},m_{2}) & = & 1 - \frac{m_{1}^{2}+m_{2}^{2}}
{m_{1}^{2}-m_{2}^{2}}\ln \frac{m_{1}}{m_{2}} -\ln m_{1} -\ln m_{2} +O(s),
\nonumber \\   
B_{0}(p;0,m) & = & 1 -2\ln m +O(s),  \nonumber \\
B_{1}(p;m_{1},m_{2}) & = & \frac{1}{2}\frac{1}{m_{2}^{2}-m_{1}^{2}}\bigg[
\frac{m_{1}^{2}+m_{2}^{2}}{2} - \frac{m_{1}^{2}m_{2}^{2}}{m_{1}^{2}-m_{2}^{2}}
\ln \frac{m_{1}^{2}}{m_{2}^{2}}\biggr] -\frac{1}{2}B_{0}(p;m_{1},m_{2}),
\nonumber \\
B_{1}(p;0,m) & = & -\frac{1}{4} + \ln m + O(s). 
\end{eqnarray}

\section*{B}

Here we prove Eq.~\ref{for1}, $\Lambda_{\phi H N} + \Lambda_{\phi \chi N} + \delta Z_{L}^{\phi N} =  0$, using Eq.~\ref{hophop}.

We note the vertex $V_{\phi \phi N}^{\gamma}$ (Fig. \ref{gll}c) is given as
\begin{eqnarray}
\label{lorentz1}
V_{\phi \phi N}^{\gamma} & \equiv & i e \gamma_{\mu} F_{V}^{\phi \phi N}
- i e \gamma_{\mu} \gamma_{5} F_{A}^{\phi \phi N} \nonumber \\
& = & 
\sum_{a} \int \frac{d^{n}q}{(2\pi)^{n}}
\frac{+ig_{2}}{\sqrt{2}M_{W}} \big(K_{H}
\big)_{la}M_{N}\frac{1+\gamma_{5}}{2}\frac{i}{\not q - \not p_{1} -
M_{N}}\nonumber \\
& \times &           
\frac{+ig_{2}}{\sqrt{2}M_{W}} \big(K_{H}^{\dagger}\big)_{al}
M_{N}\frac{1-\gamma_{5}}{2}\frac{i}{(q-p_{1}-p_{2})^{2}-M_{W}^{2}}
\frac{i}{q^{2}-M_{W}^{2}} \nonumber \\
& \times &
i e \Big(-2q + p_{1} + p_{2}\Big)_{\mu}.
\end{eqnarray}
If we now return from the $\gamma l l$ vertex to the $W \mu \nu$ vertex (Fig.~\ref{vermuon}f), we
have
\begin{eqnarray}
\label{lorentz2}
V_{\phi H N}^{W} & = &
\sum_{a} \int \frac{d^{n}q}{(2\pi)^{n}}
\frac{-ig_{2}}{2 M_{W}}M_{N}\big(K_{L}^{\dagger}K_{H}\big)_{ia}
\frac{1+\gamma_{5}}{2}\frac{i}{\not q - \not p_{1} -
M_{N}}\nonumber \\
& \times &
\frac{+ig_{2}}{\sqrt{2}M_{W}}\big(K_{H}^{\dagger}\big)_{al} M_{N}
\frac{1-\gamma_{5}}{2} \frac{i}{(q-p_{1}-p_{2})^{2}-M_{H}^{2}} 
\frac{i}{q^{2}-M_{W}^{2}}\nonumber \\
& \times &
 \frac{i g_{2}}{2}\Big(+2q - p_{1} - p_{2}\Big)_{\mu}.
\end{eqnarray}
A similar expression holds for $V_{\phi \chi N}^{W}$.
The Lorentz structure of Eqs.~\ref{lorentz1} and \ref{lorentz2} is the same.
In the large $M_{N}$ limit, $M_{H}$ and $M_{W}$ in the propagators are
negligible; therefore the only possible difference between the two vertices
comes from constant factors. If we forget for a moment about the mixing factors,
it can be easily checked that 
\begin{eqnarray}
\label{hoptrop}
V_{\phi H N}^{W} & = & \frac{\sqrt{2} g_{2}}{4 e}V_{\phi \phi N}^{\gamma}, \nonumber \\ 
{\cal M}_{\phi H N} + {\cal M}_{\phi \chi N} & \equiv & 
(\Lambda_{\phi H N} + \Lambda_{\phi \chi N}){\cal M}_{tree} \nonumber
\\
& = & 
\overline{u}_{\nu_{\mu}}(V_{\phi H N}^{W} + V_{\phi \chi N}^{W})u_{\mu}
\frac{i g^{\mu\nu}}{M_{W}^{2}} \overline{v}_{e} \frac{i g_{2}}{2\sqrt{2}}
\big(K_{L}\big)_{ej}\gamma_{\nu}(1-\gamma_{5}) v_{\nu_{e}},
\nonumber \\
\Lambda_{\phi H N} + \Lambda_{\phi \chi N}& = & 
2 F_{V}^{\phi \phi N} \; = 2 F_{A}^{\phi \phi N}  .
\end{eqnarray}
Hence (using Eq.~\ref{hoptrop} and Eq.~\ref{hophop})
\begin{eqnarray}
\label{wardw}
\Lambda_{\phi H N} + \Lambda_{\phi \chi N} + \delta Z_{L}^{\phi N} & = &
2 F_{V}^{\phi \phi N} + \delta Z_{V}^{\phi N} + \delta Z_{A}^{\phi N} 
\nonumber \\
& = &
2 ( F_{V}^{\phi \phi N} + \delta Z_{V}^{\phi N} ) \; = \; 0.
\end{eqnarray}
That is, the two dominant nonstandard contributions from Eq.~\ref{eq9} cancel.
% This implies that the vertices can be reliably represented by the SM
%terms.
To show that the inclusion of the mixing factors will not affect 
Eq.~\ref{wardw}, note that 
the mixing factor for the $W\mu\nu$ vertex, which we denote as $k_{1}$,  is 
related to that of the
$\gamma ll$ vertex, denoted as $k_{2}$, in the approximation in which
flavour-violating mixing  factors
$e\mu_{mix}, \tau\mu_{mix}, e\tau_{mix}$ are vanishing, as follows.
\begin{eqnarray}
k_{1} & = & \sum_{a} \big(K_{H}^{\dagger}\big)_{a \mu}
\big(K_{L}^{\dagger}K_{H}\big)_{ia} \nonumber \\
& = & \big(K_{L}^{\dagger}\big)_{ie} \sum_{a} \big(K_{H}\big)_{ea}
\big(K_{H}^{\dagger}\big)_{a \mu} 
+ \big(K_{L}^{\dagger}\big)_{i\mu} \sum_{a} \big(K_{H}\big)_{\mu a}
\big(K_{H}^{\dagger}\big)_{a \mu}
+ \big(K_{L}^{\dagger}\big)_{i\tau} \sum_{a} \big(K_{H}\big)_{\tau a}
\big(K_{H}^{\dagger}\big)_{a \mu} \nonumber \\
& = & 
\big(K_{L}^{\dagger}\big)_{ie} e\mu_{mix} +
\big(K_{L}^{\dagger}\big)_{i\mu} \mu\mu_{mix} +
\big(K_{L}^{\dagger}\big)_{i\tau} \tau\mu_{mix} \nonumber \\
& = &
\big(K_{L}^{\dagger}\big)_{i\mu} \mu\mu_{mix} \; = \;
\big(K_{L}^{\dagger}\big)_{i\mu} k_{2} 
\end{eqnarray}
The remaining factor, $\big(K_{L}^{\dagger}\big)_{i\mu}$,
 is absorbed into ${\cal
M}_{tree}$ as required.

%%%%%%%%%%%%%%%%%%%%%%%%%%REFERENCES%%%%%%%%%%%%%%%%%%%%%%%%%%%

%%%%%%%%%%%%%%%%%%%%%%%%FIGURE CAPTIONS%%%%%%%%%%%%%%%%%%%%%%%%%%%
\begin{figure}[h]
\caption[]{Box diagrams for muon decay.           
\label{boxmuon}}
\end{figure}

\begin{figure}
\caption[]{Vertex diagrams for muon decay.
\label{vermuon}}
\end{figure}

\begin{figure}
\caption[]{Charged lepton self-energies.
\label{selffd}}
\end{figure}

\begin{figure}
\caption[]{$\gamma l l$ vertex .
\label{gll}}
\end{figure}

\begin{figure}
\caption[]{Neutrino self-energy diagrams for muon decay.
\label{selfmuon}}
\end{figure}

\begin{figure}
\caption[]{Counterterm diagram for neutrino self-energy in muon decay.
\label{countl}}
\end{figure}

\begin{figure}
\caption[]{Z leptonic width as a function of
$M_{N}$ for $ee_{mix}=0$, 
$m_{t}=176$ GeV, Higgs mass $= 200$ GeV and different values of the mixing parameter $\tau\tau_{mix}$,
(a) $Z \rightarrow \tau\tau$ mode,
(b) $Z \rightarrow ee$ mode.
The dashed lines
represent the $1 \sigma$ band about the current experimental value
(a)$\Gamma_{\tau\tau}^{exp} = 83.85 \pm 0.29$ MeV,
(b)$\Gamma_{ee}^{exp} = 83.92 \pm 0.17$~MeV.}
\label{resultsa}
\end{figure}

\begin{figure}
\caption[]{Z leptonic width as a function of
$M_{N}$ for $ee_{mix}=0.0071$,
$m_{t}=176$ GeV, Higgs mass $= 200$ GeV and different values of the mixing parameter $\tau\tau_{mix}$,
(a) $Z \rightarrow \tau\tau$ mode,
(b) $Z \rightarrow ee$ mode.
The dashed lines
represent the $1 \sigma$ band about the current experimental value
(a)$\Gamma_{\tau\tau}^{exp} = 83.85 \pm 0.29$ MeV,
(b)$\Gamma_{ee}^{exp} = 83.92 \pm 0.17$~MeV.}
\label{resultsb}
\end{figure}

\begin{figure}
\caption[]{The branching ratio $Z \rightarrow l_{1}^{\pm}l_{2}^{\mp}$
as a function of $M_{N}$ for (a) $\delta = -1$, (b) $\delta = +1$.}
\label{fviolfig}
\end{figure}

\newpage
\begin{table}[htb]
\begin{center}
\begin{tabular}{|l|r|r|r|r|r|r|} \hline
                  & SM  & $M_{N}$ = 0.5 TeV  & $M_{N}$ = 5 TeV & $M_{N}$ = 15 TeV & $M_{N}$ = 30 TeV &            \\
  \hline
\multicolumn{7}{|c|}{$ee_{mix} = 0.0071,\; \mu\mu_{mix} = 0.0014,\; \tau\tau_{mix} = 0.0$}\\ \hline
${\hat \Sigma}^{\nu_{e}}+ {\hat \Sigma}^{\nu_{\mu}}$  
      & - 4.995  & - 4.972$\;\;\;\;\;$  & - 4.982$\;\;\;\;\;$ & - 4.988$\;\;\;\;\;$ & - 4.992$\;\;\;\;\;$ & $\times 10^{-2}$ \\
${\hat \Lambda}^{\mu}$
      & - 1.441  & - 1.442$\;\;\;\;\;$ &  - 1.444$\;\;\;\;\;$ & - 1.444$\;\;\;\;\;$ & -1.445$\;\;\;\;\;$  & $\times 10^{-2}$ \\
${\cal M}_{box}/{\cal M}_{tree}$
      &   4.273  &   4.300$\;\;\;\;\;$  &   4.315$\;\;\;\;\;$ &   4.457$\;\;\;\;\;$ &  4.950$\;\;\;\;\;$  & $\times 10^{-3}$ \\
$\delta_{V}$
      &   6.670  &   6.539$\;\;\;\;\;$  &   6.525$\;\;\;\;\;$ &   6.652$\;\;\;\;\;$ &  7.133$\;\;\;\;\;$  & $\times 10^{-3}$ \\
${\hat \Sigma}_{W}(0)/M_{W}^{2}$
      &   2.396  &   2.346$\;\;\;\;\;$  &   2.301$\;\;\;\;\;$ &   1.872$\;\;\;\;\;$ &  0.329$\;\;\;\;\;$  & $\times 10^{-2}$ \\
$\Delta r$
      &   3.063  &   3.000$\;\;\;\;\;$  &   2.954$\;\;\;\;\;$ &   2.537$\;\;\;\;\;$ &  1.043$\;\;\;\;\;$  & $\times 10^{-2}$ \\
$M_{W}$ [GeV]           
      &  80.459  &  80.537$\;\;\;\;\;$  &  80.545$\;\;\;\;\;$ &  80.612$\;\;\;\;\;$ & 80.846$\;\;\;\;\;$  & $\times 1$       \\ 
\hline 
\multicolumn{7}{|c|}{$ee_{mix} = 0.0071,\; \mu\mu_{mix} = 0.0014,\; \tau\tau_{mix} = 0.033$}\\ \hline
$\delta_{V}$
      &   6.670  &   6.538$\;\;\;\;\;$  &   6.627$\;\;\;\;\;$ &   8.050$\;\;\;\;\;$ &  ---$\;\;\;\;\;$  & $\times 10^{-3}$ \\
${\hat \Sigma}_{W}(0)/M_{W}^{2}$
      &   2.396  &   2.363$\;\;\;\;\;$  &   1.209$\;\;\;\;\;$ &  - 13.322$\;\;\;\;\;$ &  ---$\;\;\;\;\;$  & $\times 10^{-2}$ \\
$\Delta r$
      &   3.063  &   3.017$\;\;\;\;\;$  &   1.871$\;\;\;\;\;$ &  - 12.517$\;\;\;\;\;$ &  ---$\;\;\;\;\;$  & $\times 10^{-2}$ \\
$M_{W}$ [GeV]           
      &  80.459  &  80.534$\;\;\;\;\;$  &  80.718$\;\;\;\;\;$ &    82.549$\;\;\;\;\;$ &  ---$\;\;\;\;\;$  & $\times 1$       \\ 
\hline
\end{tabular}
\end{center}
\caption{Contribution of the muon decay loops to $\delta_{V}$, $\Delta_{r}$ and
$M_{W}$.}
\label{muonloops}
\end{table}                             

\begin{table}[htb]
\begin{center}
\begin{tabular}{|c|c|c|c|}\hline
          Process  & Experimental BR \cite{pdb} & Theoretical BR & Limits on masses and/or mixings    \\
  \hline
$\mu \rightarrow e \gamma$  
      & $\leq 4.9 \times 10^{-11}\;\; 90 \%$ C.L.  & $4.9 \times 10^{-11}\;$ ${}^{a}$   & $|e\mu_{mix}| \leq 0.00024$ \\
$\tau \rightarrow e \gamma$
      & $\leq 1.2 \times 10^{-4}\;\; 90 \%$ C.L.  & $7 \times 10^{-7}\;$ ${}^{b}$  &  ---   \\
$\tau \rightarrow \mu \gamma$
      & $\leq 4.2 \times 10^{-6}\;\; 90 \%$ C.L.  & $7 \times 10^{-7}\;$ ${}^{b}$  &  ---  \\
$\mu \rightarrow e e e$
      & $\leq 1.0 \times 10^{-12}\;\; 90 \%$ C.L.   & $1.0 \times 10^{-12}\;$ ${}^{c}$   & $M_{N}^{2} \leq 0.93 \times 10^{-5} \frac{1 TeV^{2}}{ee_{mix}|e\mu_{mix}|}$ \\
$\tau \rightarrow e e e$
      & $\leq 1.4 \times 10^{-5}\;\; 90 \%$ C.L.     & $ 5 \times 10^{-7}\;$ ${}^{d}$  &  ---   \\
$\tau \rightarrow e \mu \mu$
      & $\leq 1.4 \times 10^{-5}\;\; 90 \%$ C.L.   &  $3 \times 10^{-7}\;$ ${}^{d}$  & --- \\
$Z \rightarrow e \mu$           
      & $\leq 6.0 \times 10^{-6}\;\; 95 \%$ C.L.   & $3.3 \times 10^{-8}\;$ ${}^{e}$ & ---  \\ 
$Z \rightarrow e \tau$           
      & $\leq 1.3 \times 10^{-5}\;\; 95 \%$ C.L.   & $1.4 \times 10^{-6}\;$ ${}^{e}$ & ---   \\ 
$Z \rightarrow \mu \tau$           
      & $\leq 1.9 \times 10^{-5}\;\; 95 \%$ C.L.   & $2.2 \times 10^{-7}\;$ ${}^{e}$ & ---   \\
\hline \hline
\multicolumn{4}{|l|}{${}^{a}$ $|e\mu_{mix}| = 0.00024, \; M_{N} > 0.5$ TeV; $\;\;\;\;\;$ Ref. \cite{ggjv}}\\ \hline
\multicolumn{4}{|l|}{${}^{b}$ $ee_{mix} = 0.043, \mu\mu_{mix} = 0.008, \tau\tau_{mix} = 0.1, \; M_{N} > 0.5$ TeV; $\;\;\;\;\;$ Ref. \cite{ggjv}}\\ \hline
\multicolumn{4}{|l|}{${}^{c}$ $M_{N}^{2} \times ee_{mix}|e\mu_{mix}| = 0.93 \times 10^{-5} \times 1\;$TeV$^{2};\;\;\;\;\;$ Ref. \cite{Jarlskog}}\\ \hline
\multicolumn{4}{|l|}{${}^{d}$ $ee_{mix} = 0.01, \mu\mu_{mix} = 0, \tau\tau_{mix} = 0.033, M_{N} = 3\;$ TeV; $\;\;\;\;\;$ Ref. \cite{pilaftsis1} }\\ \hline
\multicolumn{4}{|l|}{${}^{e}$ $ee_{mix} = 0.0071, \mu\mu_{mix} = 0.0014, \tau\tau_{mix} = 0.033, M_{N} = 5\;$TeV; $\;\;\;\;\;$ this paper and Ref. \cite{pilaftsis1}}\\ 
\hline
\end{tabular}     
\end{center}
\caption{Flavour-violating decays: experimental limits, theoretical predictions and the constraints implied.}
\label{tabulka}
\end{table}

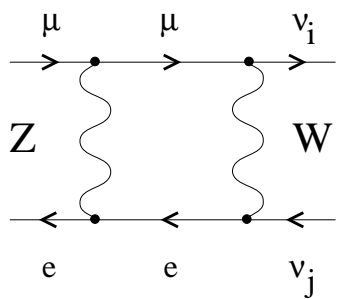
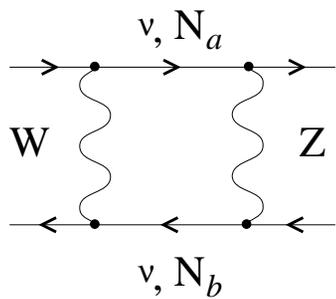
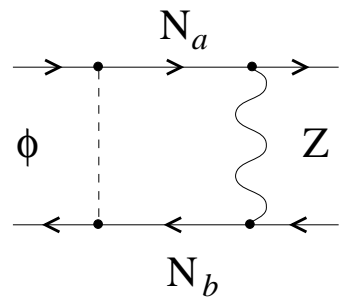
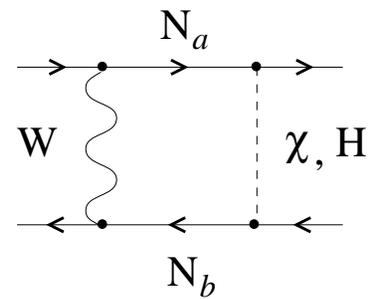
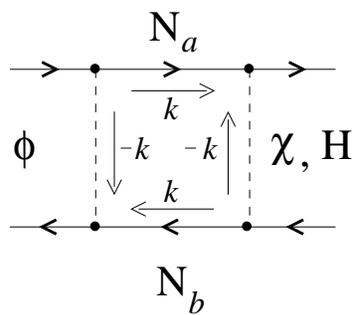
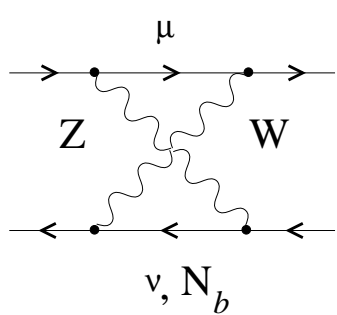
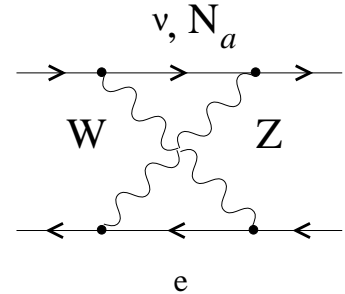
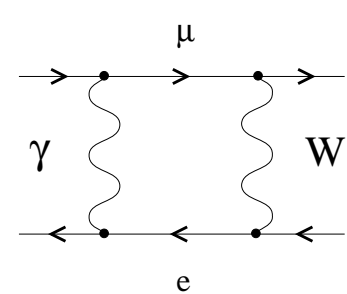
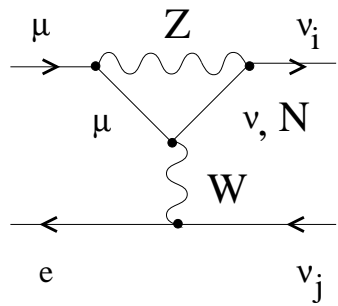
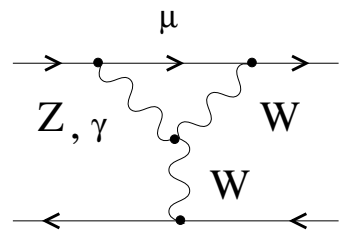
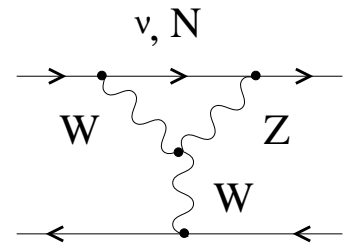
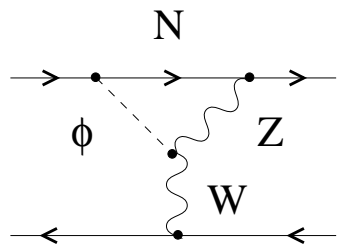
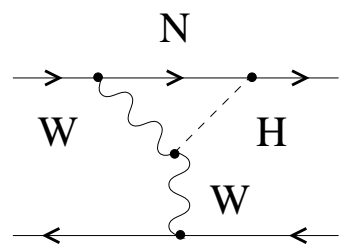
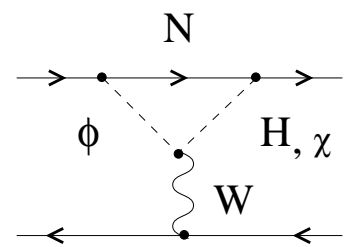
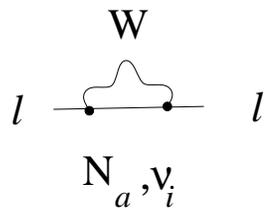
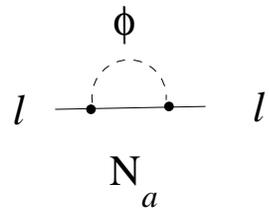
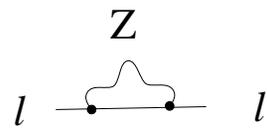
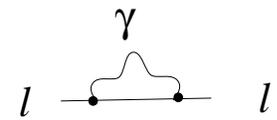
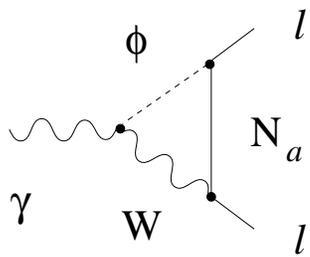
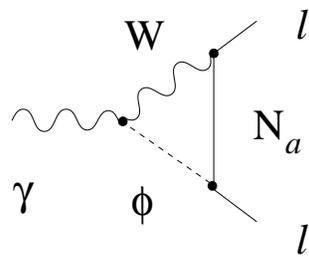
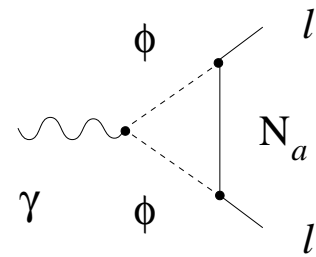
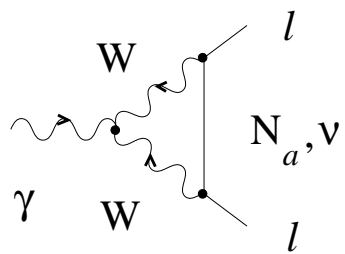
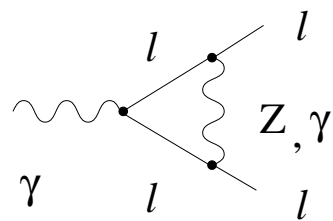
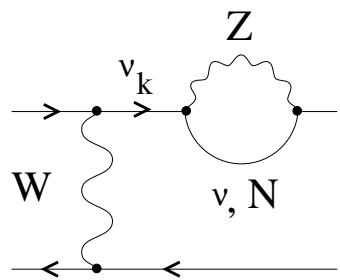
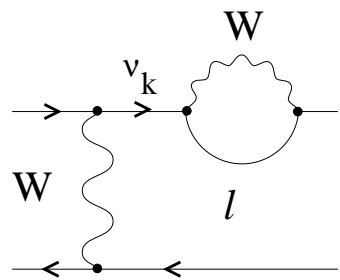
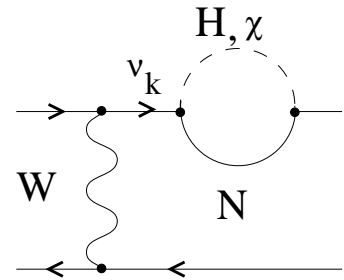
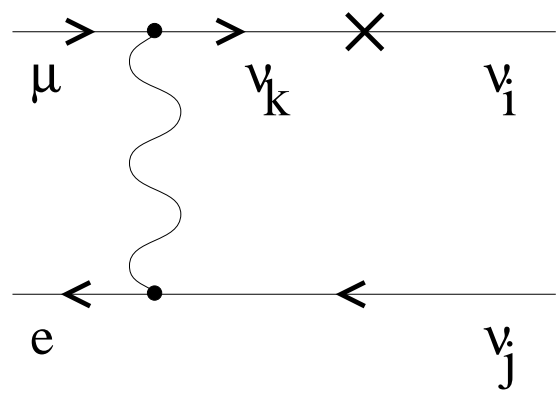
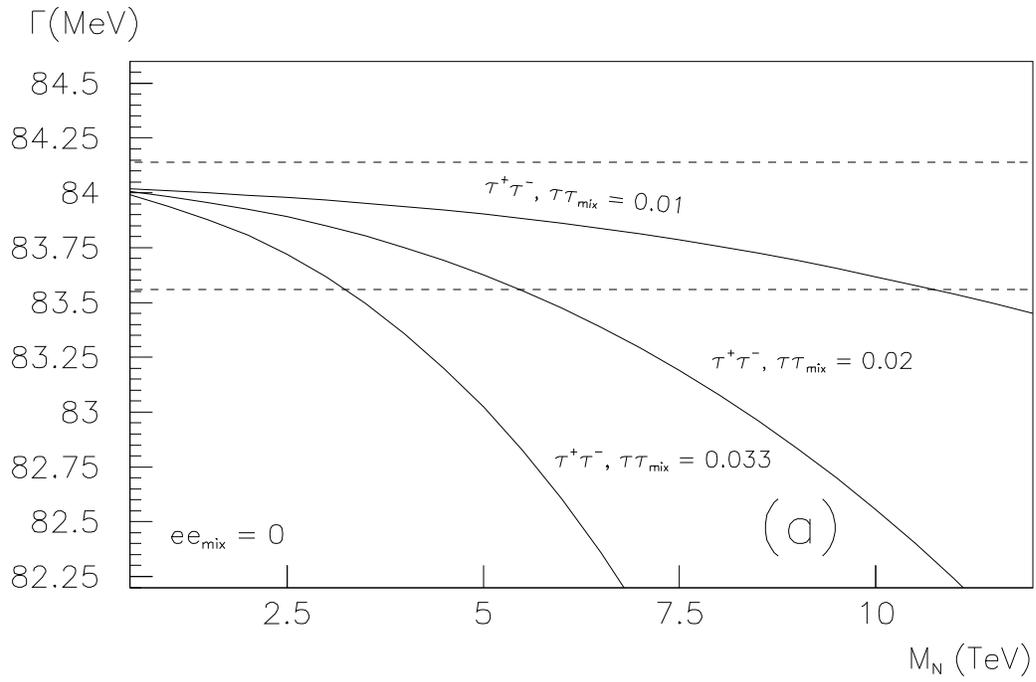
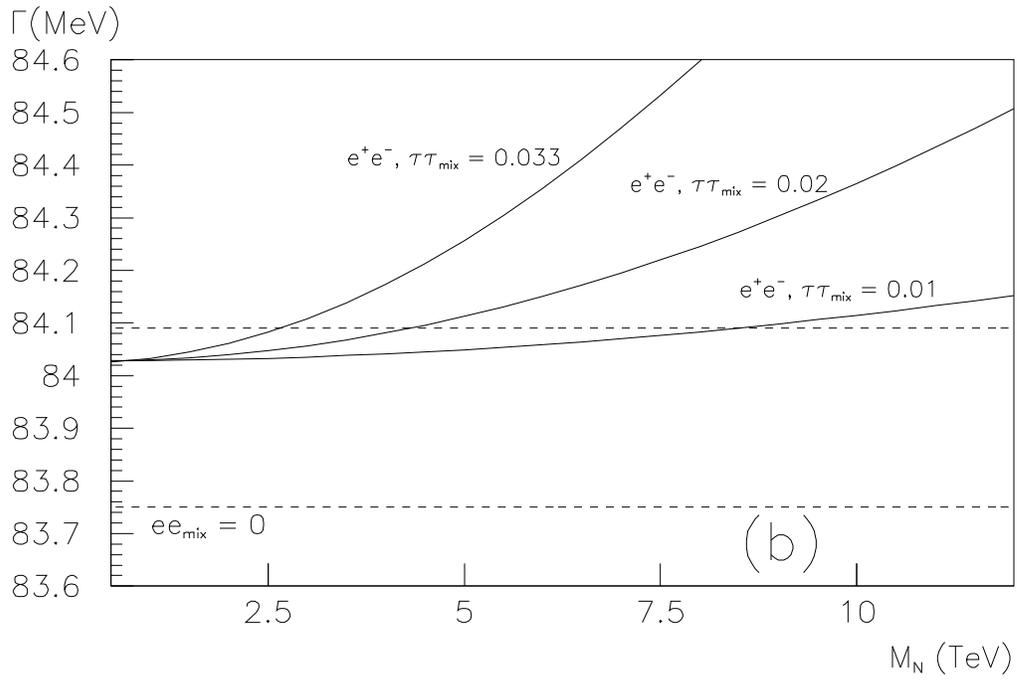
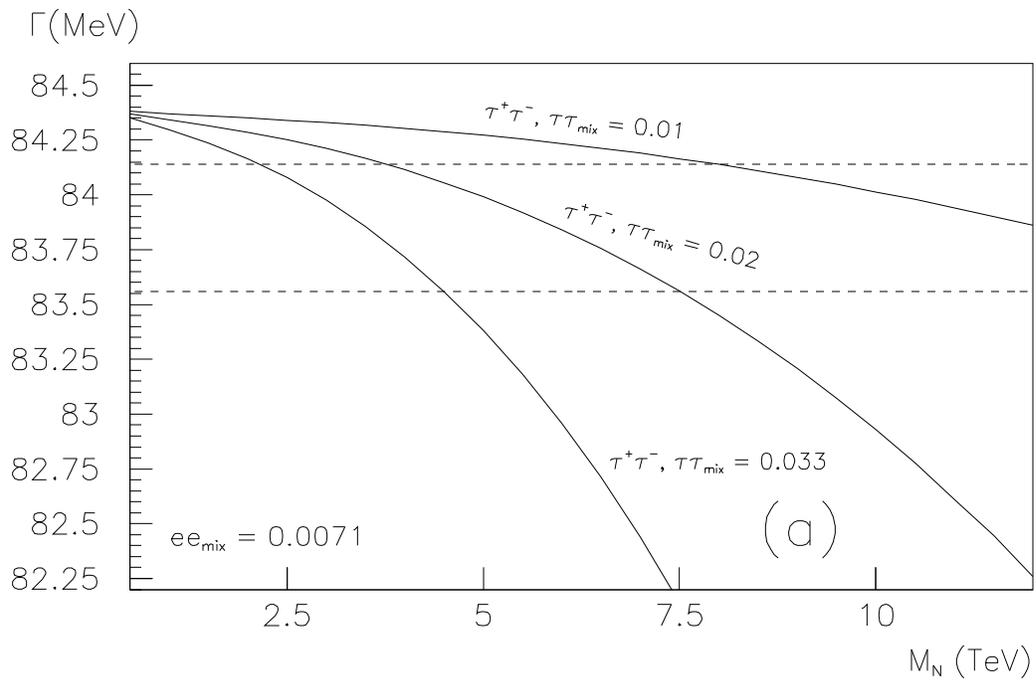
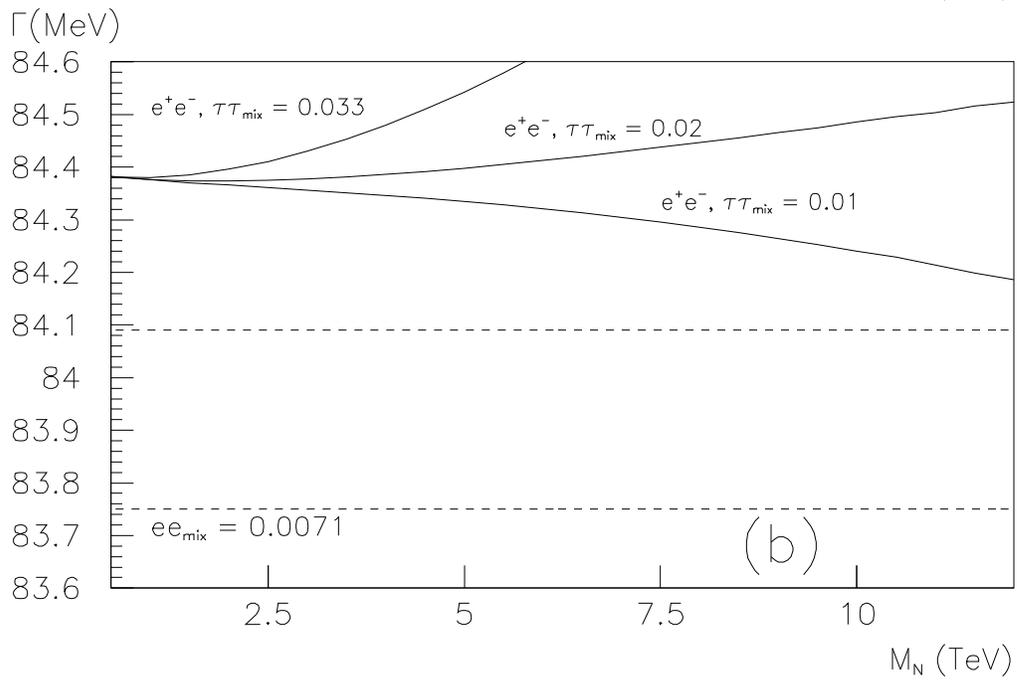
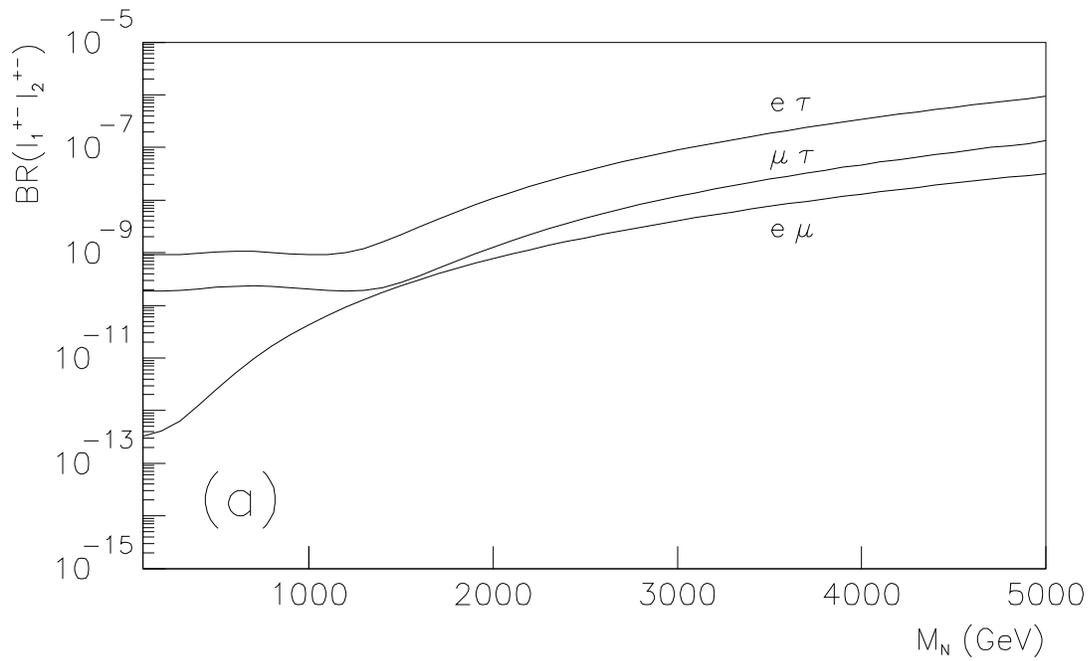
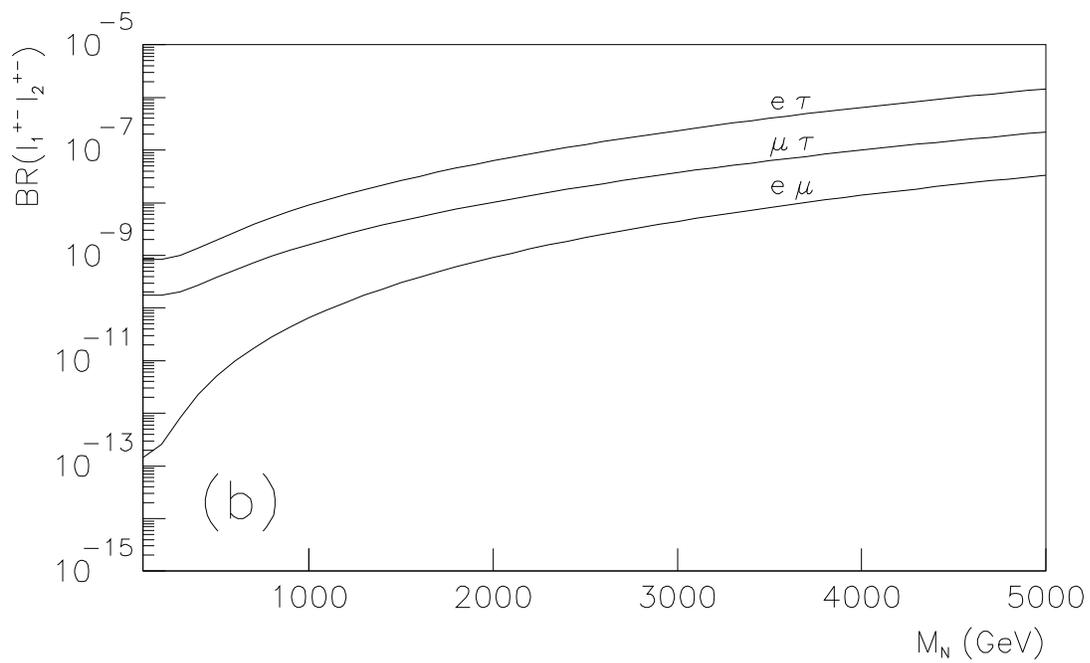
\end{document}